\UseRawInputEncoding

\documentclass[aps,prl,twocolumn,amsmath,amssymb,nofootinbib,superscriptaddress,floatfix]{revtex4-1}

\usepackage{xcolor}
\usepackage{gensymb}
\usepackage{amsmath}
\usepackage{amssymb}
\usepackage{amsthm}
\usepackage{mathtools}
\usepackage{mathdots}
\usepackage{amsfonts}
\usepackage{bbm}
\usepackage{xr}
\usepackage{bbold}
\usepackage{newpxtext}
\usepackage{epsfig,color,dsfont,upgreek,physics}
\usepackage{mathrsfs}
\usepackage{multirow}
\usepackage[makeroom]{cancel}
\usepackage{mathtools}

\usepackage{graphicx}
\usepackage{dcolumn} 
\usepackage{bm} 
\usepackage{float} 
\usepackage[driverfallback=dvipdfm]{hyperref} 
\usepackage{longtable}
\usepackage{ulem}   
\allowdisplaybreaks

\newcommand{\bs}[1]{{\boldsymbol{#1}}}

\begin{document}

 \title{
Fractional Chern Insulators in Twisted Bilayer MoTe$_2$: A Composite Fermion Perspective
}

\author{Tianhong Lu}
\affiliation
{
Department  of  Physics,  Emory  University,  400 Dowman Drive, Atlanta,  GA  30322,  USA
}
\author{Luiz H. Santos}
\affiliation
{
Department  of  Physics,  Emory  University,  400 Dowman Drive, Atlanta,  GA  30322,  USA
}

\begin{abstract}
The discovery of Fractional Chern Insulators (FCIs) in twisted bilayer MoTe$_2$ has sparked significant interest in fractional topological matter without external magnetic fields. Unlike the flat dispersion of Landau levels, moir\'e electronic states are influenced by lattice effects within a nanometer-scale superlattice. This study examines the impact of these lattice effects on the topological phases in twisted bilayer MoTe$_2$, uncovering a family of FCIs with Abelian anyonic quasiparticles.
Using a composite fermion approach, we identify a sequence of FCIs with fractional Hall conductivities $\sigma_{xy} = \frac{C}{2C + 1} \frac{e^2}{h}$ linked to partial filling $\nu_{\,\text{h}}$ of holes of the topmost moir\'e valence band. These states emerge from incompressible composite fermion bands of Chern number $C$ within a complex Hofstadter spectrum. This approach explains FCIs with Hall conductivities $\sigma_{xy} = (2/3) e^2/h$ and $\sigma_{xy} = (3/5) e^2/h$ at fractional fillings $\nu_{\,\text{h}} = 2/3$ and $\nu_{\,\text{h}} = 3/5$ observed in experiments, and uncovers other fractal FCI states.
The Hofstadter spectrum reveals new phenomena, distinct from Landau levels, including a higher-order Van Hove singularity (HOVHS) at half-filling, leading to novel quantum phase transitions. This work offers a comprehensive framework for understanding FCIs in transition metal dichalcogenide moir\'e systems and highlights mechanisms for topological quantum criticality.
\end{abstract}

\date{\today}

\maketitle

\noindent
\textit{Introduction--}
The experimental discovery of Fractional Chern Insulators (FCIs) \cite{Neupert-2011,Sheng-2011,Tang-2011,Sun-2011,Regnault2011} in twisted bilayer MoTe$_2$ (tMoTe$_2$) \cite{cai2023signatures,zeng2023thermodynamic, park2023observation, xu2023observation} 
has sparked significant interest in fractional topological matter without external magnetic fields in moir\'e transition metal dichalcogenide (TMD)
systems.\cite{li2021spontaneous,crepel2023anomalous,wang2024fractional,reddy2023fractional,jia2024moire} 
The interplay of spin-orbit interaction, moir\'e layer potential, and interlayer tunneling in tMoTe$2$ produce moir\'e valence bands with spin/valley-Chern numbers \cite{wu2019topological,wang2024fractional,reddy2023fractional,jia2024moire}, providing a platform for realizing
FCIs with Hall conductivities $\sigma_{xy} = \nu_{\,\text{h}}e^{2}/h$ at hole filling $\nu_{\, \text{h}} = 2/3$ and $\nu_{\,\text{h}} = 3/5$ \cite{cai2023signatures,zeng2023thermodynamic, park2023observation, xu2023observation} when spontaneously broken time-reversal symmetry 
produces partially filled spin-polarized Chern bands.

Although FCIs exhibit the same 
transport properties as fractional quantum Hall states created by 
magnetic fields \cite{halperin2020fractional}, the microscopic origins of moir\'e bands and Landau levels are profoundly different. Landau levels exhibit flat dispersion because the magnetic length is significantly larger (nanometer scale) compared to the atomic separation (Angstrom scale) in conventional lattices. 
In contrast, moir\'e electronic states, formed by restructuring within a nanometer-scale superlattice\cite{andrei2021marvels}, are heavily influenced by lattice effects. A key question in exploring FCIs in TMD moir\'e systems is how these lattice effects impact topological phases with fractional quasiparticles.

In this study, we uncover a family of FCIs supporting Abelian anyons in partially filled valence bands of tMoTe$2$ and demonstrate  
how quantum phase transitions (QPTs) induced by lattice effects can change topological order.
Employing a composite fermion (CF) approach \cite{jain1989composite,lopez1991fractional,Kol1993,Moller2015,murthyshankar2012,Sohal-2018,wang_classification_2020} characterized by binding of two flux quanta of the Chern-Simons gauge field to electrons residing on a honeycomb lattice \cite{Haldane1988} describing the spin/valley polarized Chern bands close to charge neutrality\cite{wu2019topological,wang2024fractional,reddy2023fractional,jia2024moire}, 
we delineate a series of incompressible topological states at fractional filling $0 < \nu_{\, \text{h}} < 1$ of holes on the topmost spin-polarized valence bands of tMoTe$_2$.
The flux attachment performed on the effective Haldane-type lattice \cite{Haldane1988,wu2019topological} yields a rich fractal Hofstadter spectrum\cite{Hofstadter76} 
describing CFs coupled to a large 
dynamical Chern-Simons flux within each unit cell. 
The emergent Hofstadter spectrum, 
a core aspect of our analysis, enables us to pinpoint emergent incompressible states, predict their fractional conductivities, and, notably, uncover new quantum critical phenomena that underscore key differences between CFs in Chern bands and Landau levels.

The main results of this work are:
\\
\noindent
(1) We chart the sequence of FCIs with fractional Hall conductivity $\sigma_{xy} = \frac{C}{2C + 1}\frac{e^2}{h}$, which originate from gapped CF bands with integer Chern number $C$. 
Analysis of the CF spectra reveals that the most robust incompressible states are those \textit{Jain-FCI} states with fractional hole filling $\nu_{\, \text{h}}  = \frac{C}{2C + 1}$, which fit the pattern of hierarchical Jain states \cite{jain1989composite,lopez1991fractional}.
Among these Jain-FCIs, 
we identify states at $\nu_{\, \text{h}} =2/3$ and $\nu_{\, \text{h}} =3/5$ consistent with recent experimental observations \cite{cai2023signatures,zeng2023thermodynamic, park2023observation, xu2023observation}. 

While the $2/3$ and $3/5$ states in tMoTe$_2$ were studied using a continuum CF approach\cite{goldman2023zero},
we identify
other 
FCIs that deviate from the usual
hierarchy, such as those at filling $\nu_{\, \text{h}}  = \frac{1}{5},\frac{2}{7},\frac{7}{9},\frac{4}{5}$, 
characterized by
$\sigma_{xy}/(e^2/h) = \frac{3}{5},\frac{3}{7},\frac{5}{9},\frac{3}{5}$. Breaking away from the Landau-level paradigm, these \textit{fractal-FCI} states develop from small gaps in the fractal spectrum, and their experimental observation would provide a signature of fractal composite fermions.
\\
\noindent
{
(2) The principal Jain-FCI staircase converges to a compressible CF state at half-filling ($\nu_{\, \text{h}} =1/2$)
characterized by a band with parabolic dispersion with effective mass $m^{*}$. We analytically determine how $m^{*}$ depends on the lattice parameters that define the effective tMoTe$_2$ bands and, unexpectedly, discover a higher-order Van Hove singularity (HOVHS) \cite{shtyk2017electrons,yuan2019magic} linked to the divergence of $m^{*}$.
This phenomenon is distinct from the behaviors observed in half-filled Landau levels \cite{halperin1993theory,son2015composite} and other composite Fermi liquid approaches in tMoTe$_2$ \cite{goldman2023zero,dong2023composite}.
Remarkably, the proximity to this lattice-driven HOVHS 
leads to the rapid collapse of the 
CF gaps near half-filling, unveiling new mechanisms for QPTs
between FCIs, which is one of the main results of this work.
Leveraging this mechanism, we uncover several QPTs 
controlled by the hopping amplitudes.
Thus, the CF predictions offer a guiding principle for experimental and numerical exploration of novel FCIs, including
novel mechanisms for topological quantum criticality in
TMD systems.
}

\begin{figure}[!t]
    \centering
    \includegraphics[width = 9 cm]{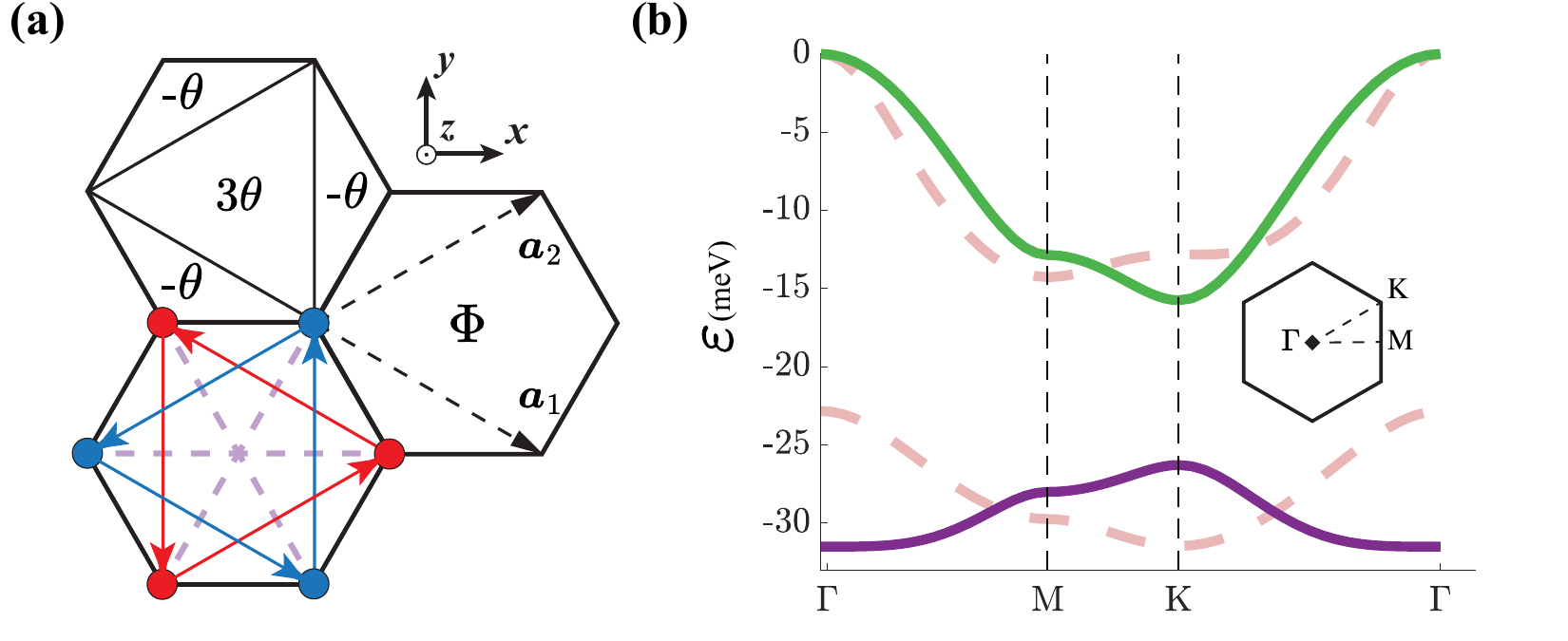}
\caption{Effective lattice and Chern bands of tMoTe$_2$.
(a) Honeycomb lattice with sublattices $A$ ($B$) in red (blue). The arrows denote the second-neighbor hopping $e^{i\theta_{\sigma}^{\,e}}$ in Eq.\,\eqref{eq: full TB model}. Time-reversal symmetry flips spin and arrow directions. Haldane $(\theta)$ and Chern-Simons fluxes $(\Phi = \Phi_0 \cdot p/q)$ along $+z$ direction are shown; (b) Energy bands of Hamiltonian \eqref{eq: full TB model} after particle-hole transformation \eqref{eq: PH transformation} against 3.89$\degree$ DFT bands(dashed)\cite{wang2024fractional}
with 
$\theta_{\sigma}^{\,h}=-2\pi/3$, $t_0,t_1,t_2=-5.83,-1.17,0.58$ (meV): the lower (upper) band carries Chern number $C=1$($-1$); band gap $\Delta E=10.53$ (meV).
}
\label{fig: haldanemodel}
\end{figure}

\noindent
\textit{Composite Fermions in tMoTe$_2$--}
For twist angle $\theta_{twist} \lesssim 4^{\degree}$, 
tMoTe$_2$ supports topmost valence bands with Chern numbers $C = \pm 1$ and spin-valley polarization $\sigma = \pm 1$\cite{wu2019topological,wang2024fractional,reddy2023fractional,jia2024moire}.
We introduce an extended Haldane-like model\cite{Haldane1988, wu2019topological} up to third-nearest neighbor hopping
with sublattices $A$ and $B$ (denoted $\ell = \pm1$), 
to describe this pair of Chern bands at ${\theta_{twist}} \approx 3.98^{\degree}$,
\begin{equation}
\label{eq: full TB model}
\begin{split}
    H_{\sigma} = 
\sum_{\substack{\ell = \pm1\\\langle\bs{r},\bs{r}'\rangle,\bs{\delta_1}}}
&-t_{0}\,c^{\dagger}_{\ell\sigma\bs{r}}\,c_{-\ell\sigma\bs{r}'}
 -t_{1}\,
e^{i\theta_{\sigma}^{\,e}}
\,c^{\dagger}_{\ell\sigma\bs{r}+\bs{\delta_1}}\,c_{\ell\sigma\bs{r}} \\
\sum_{\bs{\delta_2}}&-t_{2}\,c^{\dagger}_{\ell\sigma\bs{r}+\bs{\delta_2}}\,c_{-\ell\sigma\bs{r}} + \textrm{H.c.}
\end{split}
\end{equation}
where the Haldane phase $\theta_{\sigma}^{\,e} = -2\pi\sigma/3$ is fixed by the symmetries of the emergent hexagonal moir\'e superlattice\cite{wu2019topological,Supplemental},
$\langle,\rangle$ denotes nearest neighbor,
$\bs{\delta_1}$ and $\bs{\delta_2}$ connect, respectively, next- and third-nearest neighbors,
$c^{\dagger}_{\ell \sigma\bs{r}}$ is the electron creation operator with spin $\sigma$ at $\bs{r}$, 
and $\bs{a}_1=a\,(3/2, -\sqrt{3}/2),\bs{a}_2=a\,(3/2, \sqrt{3}/2)$ 
are the two primitive vectors (henceforth, $a=1$) 
connecting a site to its next-nearest neighbors 
(see Fig.\,\ref{fig: haldanemodel}(a)). 
To capture the phenomenology of FCIs, we focus on electrons (holes) with fixed spin polarization $\sigma = -1$ resulting from the spontaneous breaking of time-reversal symmetry due to interactions\cite{wu2019topological,wang2024fractional,reddy2023fractional,jia2024moire}. 

To characterize FCIs at hole filling $\nu_{\,\text{h}}$ of the highest valence band\cite{cai2023signatures,zeng2023thermodynamic, park2023observation, xu2023observation},
we perform a particle-hole (PH) transformation $c_{\ell\sigma}\rightarrow c^{\dagger}_{\ell\sigma}$ that maps the parameters of 
\eqref{eq: full TB model} to 
\begin{equation}
\label{eq: PH transformation}
\{-t_0,-t_1,-t_2, \theta_{\sigma}^{\,e}\} \xrightarrow{\text{PH}} \{t_0,t_1,t_2, -\theta^{\,e}_{\sigma}\}
\,,
\end{equation}
which yields a tight-binding model
for holes with hopping parameters \{$t_i$\} and phase $\theta_{\sigma}^{\,h}=-\theta_{\sigma}^{\,e}=2\pi\sigma/3$.
Fig.\,\ref{fig: haldanemodel}(b) shows the valley-polarized bands for holes with Chern numbers $C = \pm 1$, at a representative point $t_0,t_1,t_2 =-5.83, -1.17, 0.58$ (meV), or $t_0,t_1,t_2 =-1, -0.2, 0.1$  in the unit of $|t_0|$,
in the phase space, with the partial filling $0 < \nu_{\,\text{h}} < 1$ of the lowest band being the subject of this work. While direct experimental access to the energy bands of tMoTe$_2$ is not yet available, the Haldane model (\ref{eq: full TB model}) sufficiently captures both the topological properties and energy scales observed in DFT bands at a twist angle of $\theta_{twist} = 3.89^{\degree}$\cite{wang2024fractional}, offering a viable framework for studying FCI states in tMoTe$_2$.

We characterize the onset of incompressible states at partial band filling via a Chern-Simons flux attachment that maps a partially filled Chern band into a set of filled CF bands \cite{Moller2015,murthyshankar2012,Sohal-2018,wang_classification_2020}.
Restricting analysis to \textit{uniform} particle density states with two Chern-Simons flux quanta (2$\Phi_0$) per electron results in the uniform flux per unit cell $\Phi$ (see Fig.\,\ref{fig: haldanemodel}(a)) proportional to the electron lattice filling $n\in [0,1]$ (or the electron band filling $\nu_{\,\text{e}}=2n\in [0,2]$),
\begin{equation}
\label{eq:flux density relation}
    \Phi/\Phi_0 
    = 
    2\cdot 2n
    =
    2\nu_{\,\text{e}}
    =
    2\,(2-\nu_{\,\text{h}})
    \,,
\end{equation}
where 
the coefficient of $n$ in the first equality of Eq.\,\eqref{eq:flux density relation} represents the product of 2 fluxes and 2 sublattices, and the last 
equality follows from charge conservation $\nu_{\,\text{e}}+\nu_{\,\text{h}}=2$.
We emphasize that, although we analyze the onset of incompressible CF states by examining the Hofstadter spectrum of holes as a function of filling $0 < \nu_{\,\text{h}} < 1$, it is important to note that \textit{Chern-Simons flux attachment is performed on electrons, not on holes}. This distinction is crucial because flux attachment to electrons accurately captures the FCI states observed in experiments, as we will discuss further. Additionally, we have verified that the Hofstadter spectrum of electrons exhibits the same Jain sequence at the corresponding electron filling $1<\nu_{\text{e}}<2$.

The Chern-Simons 
field is incorporated into $H_{\sigma=-1}$ in Eq.\,\eqref{eq: full TB model} 
using the Peierls substitution
\begin{equation}
\label{eq: Peierls}
t_{\bs{r},\bs{r}'}\,c^{\dagger}_{\bs{r}}c_{\bs{r}'} \rightarrow
t_{\bs{r},\bs{r}'}\,
e^{i (2\pi/\Phi_0) \int^{\bs{r}}_{\bs{r}'} d \bs{l} \cdot \bs{a}}
\,
c^{\dagger}_{\bs{r}}c_{\bs{r}'}
\,,
\end{equation}
where $\bs{a}$ is the Chern-Simons vector potential 
in the gauge
$\bs{a} = \frac{\Phi}{3\sqrt{3}/2}(x+\sqrt{3}y)\,\hat{e}_y$. More details are provided in the Supplemental Materials\cite{Supplemental}.
\begin{figure}[!htb]
    \centering
    \includegraphics[width = 9 cm]{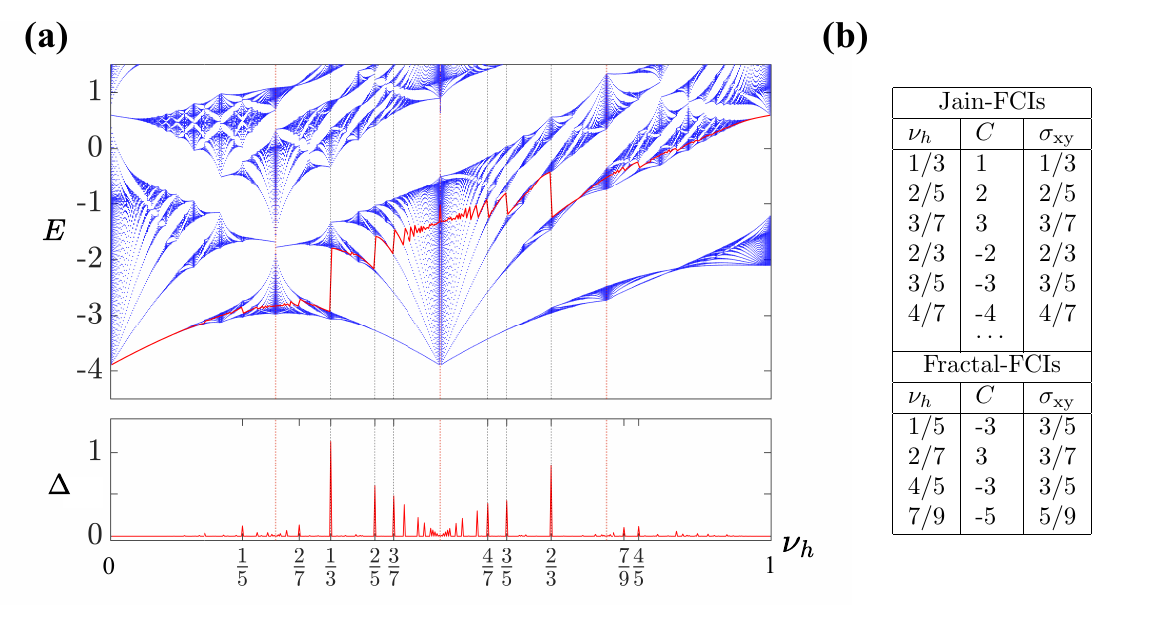}
    \caption{Composite fermion spectrum and emergent FCIs for $\theta_{\sigma}^{\,h}=-2\pi/3$, and $(t_0,t_1,t_2)=(-1,-0.2,0.1)$ in the unit of $|t_0|$.
    (a) Upper part: Hofstadter spectrum $E$ versus $\nu_{\,\text{h}}$, with the red curve showing the Fermi energy. Lower part: The composite fermion gap $\Delta$. Red dashed lines show compressible states at the $1/4,1/2,3/4$ hole fillings; (b) Fractional Hall conductance: table of Jain-FCI and fractal-FCI states at hole band filling $\nu_{\,\text{h}}$.
    } 
    \label{Fig2}
\end{figure}
For rational  
Chern-Simons flux $\Phi/\Phi_0 = p/q$ (with $p$ and $q$ coprime), the 
$C = \pm 1$ bands splits into $2q$ CF bands, resulting in the Hofstadter spectrum of CFs
shown in Fig.\,\ref{Fig2}(a).
This spectrum has $\Phi \rightarrow \Phi + 6\Phi_0$
periodicity since the
smallest triangle subunit pierced by flux shown in Fig.\,\ref{fig: haldanemodel}(a) is $1/6$ of the unit cell area.
We observe incompressible CF states, indicated by vertical jumps in the composite fermion Fermi energies (red line in Fig.\,\ref{Fig2}(a))
whenever integer bands are filled. This identifies composite fermion gaps $\Delta$ as a function of hole filling $\nu_{\,\text{h}}$.

The Chern number $C \in \mathbb{Z}$ \cite{TKNN} of the 
CF insulator is related to the Hall conductance of as\cite{Moller2015,Supplemental}
\begin{equation}
\label{eq: sigma to C relation}
    \sigma_{\text{xy}} 
    = 
    {C}/{(2C+1)}.
\end{equation}

Applying Eqs. \eqref{eq:flux density relation} and \eqref{eq: sigma to C relation} to
CF gaps shown in Fig.\,\ref{Fig2}(a), uncovers two classes of candidate FCIs in tMoTe$_2$. 
First, we identify states where $\sigma_{\text{xy}} = \nu_{\,\text{h}}\,\frac{e^2}{h}$, dubbed Jain-FCI states for the analogy to hierarchical Jain states \cite{jain1989composite,lopez1991fractional} in Landau level systems. 
Alternatively, we also identify fractal-FCI states with $\sigma_{\text{xy}} \neq \nu_{\,\text{h}}\,\frac{e^2}{h}$ 
originating from gapped fractal bands distinct from Landau levels.

\begin{figure}
    \centering
    \includegraphics[width=9cm]{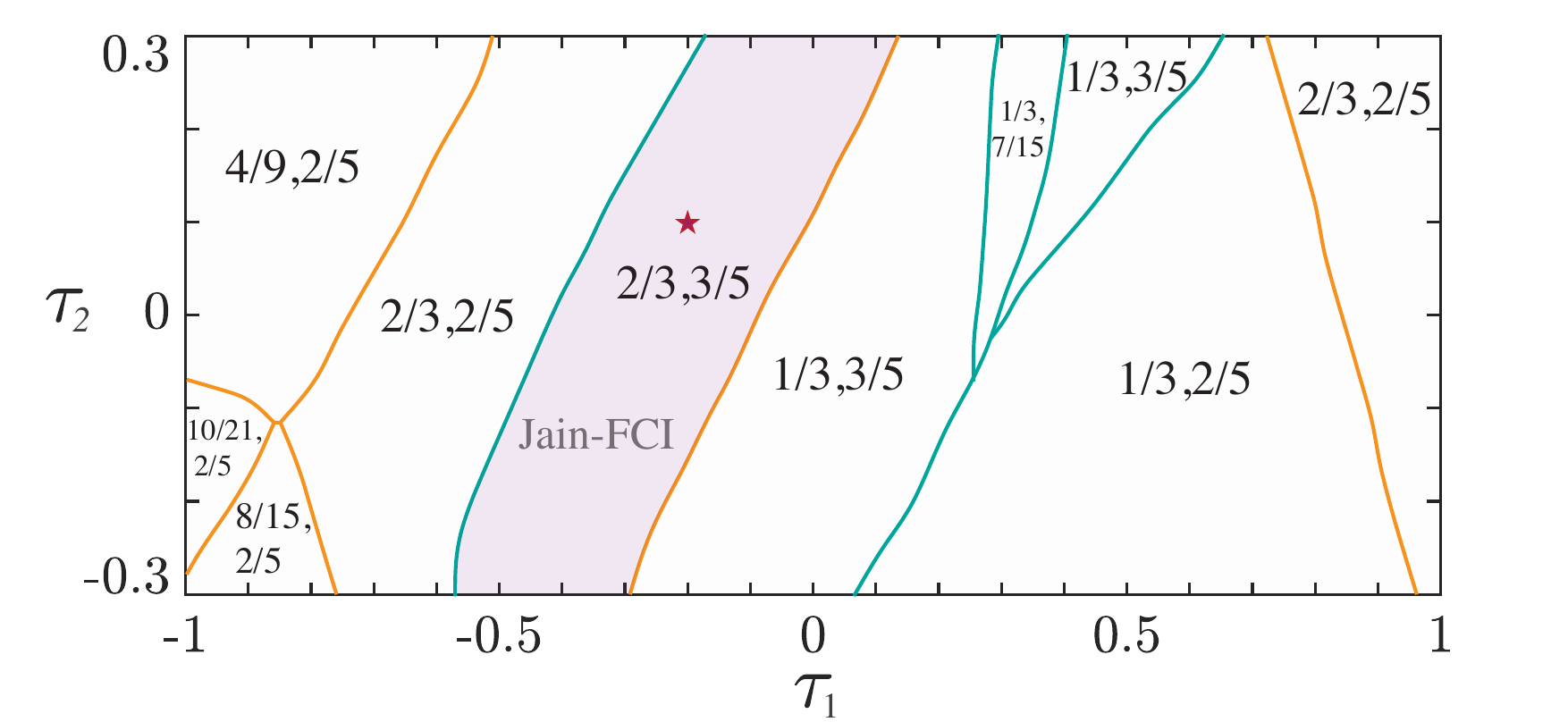}
    \caption{Phase diagram of Hall conductance $\sigma_{xy}^{2/3},\sigma_{xy}^{3/5}$ for $\nu_{\,\text{h}}=2/3,3/5$, as function of $(\tau_1,\tau_2)$. Star: $(\tau_1,\tau_2) = (-0.2, 0.1)$, the representative point in Figs.\,\ref{fig: haldanemodel} and \ref{Fig2}.
    }
    \label{fig: joint phase diagram}
\end{figure}

In Fig.\,\ref{Fig2}(a) and the upper part of Fig.\,\ref{Fig2}(b), we depict a series of Jain-FCI states, in the unit of $|t_0|$, for $t_0=-1$, $\tau_1\equiv t_1/|t_0| = -0.2$, and $\tau_2\equiv t_2/|t_0| = 0.1$.
Despite recent variations in the spectrum of Chern bands in tMoTe$_2$ observed in density functional methods \cite{wang2024fractional,reddy2023fractional,jia2024moire}, our results are robust for a range of hoppings.
The most prominent gaps occur at $\nu_{\,\text{h}} = 1/3$ and $\nu_{\,\text{h}} = 2/3$, 
with noted particle-hole asymmetry.
Then, the composite fermion gap $\Delta$ decreases starting from the $2/5,3/5$ hole filling towards $\nu_{\,\text{h}}=1/2$. \textit{Notably, we identify Jain-FCIs at $\nu_{\,\text{h}}=2/3,3/5$, with the same topological properties as those recently observed \cite{cai2023signatures,zeng2023thermodynamic, park2023observation, xu2023observation}.} 
The joint phase diagram in Fig.\ref{fig: joint phase diagram} maps the stability of the FCIs at hole fillings of $2/3$ and $3/5$, demonstrating their robustness over varying normalized hopping strengths $\tau_1$ and $\tau_2$.
The CF approach predicts a significant $\nu_{\,\text{h}}=1/3$ gap, but experimental evidence for this FCI remains elusive, possibly due to competing states.

The Jain-FCI sequence extends and converges to a composite Fermi liquid at half-filling.
Outside this region, we identify fractal-FCI states shown 
in the lower part of Fig.\,\ref{Fig2}(b).
These originate from the fractal nature of the CF spectrum and possess gaps significantly smaller than most of the Jain-FCI gaps, e.g. at $\nu_{\,\text{h}} = 2/3$ and $\nu_{\,\text{h}} = 3/5$. 
This indicates that observing fractal-FCIs may necessitate more stringent conditions, such as lower temperatures or higher sample quality.

\noindent
\textit{Quantum Phase Transitions--}
We now shift our attention to the states at half-filling ($\nu_{\,\text{h}}=1/2$), to which the Jain sequence converges. For simplicity, we focus on the line $\tau_2=0$ in Fig.\,\ref{fig: joint phase diagram}; generalization to $\tau_2\neq0$ is discussed in \cite{Supplemental}.
The Chern-Simons flux per unit cell at $\nu_{\,\text{h}}=1/2$ is $\Phi = 3\Phi_0$ (see Eq.\ref{eq:flux density relation}), and to gain generality, we treat the Haldane phase as an independent variable $\theta$.
Then the compressible CF states are described by the Hamiltonian $H_{\nu_{\,\text{h}} = 1/2} = h_{0}(\bs{k})\sigma_0 + \sum^{3}_{i=1}h_{i}(\bs{k})\sigma_i$, 
with
\begin{equation}
\begin{split}
&\,
h_{0}(\bs{k}) = 2t_1\cos\theta\left(\cos k_1-\cos k_2+\cos(k_1-k_2)\right)
\,,
\\
&\,
h_{1}(\bs{k}) = t_0(1-\cos k_1 +\cos k_2)
\,,
\\
&\,
h_{2}(\bs{k})=t_0(-\sin k_1 +\sin k_2)
\,,
\\
&\,
h_{3}(\bs{k}) = 2t_1\sin\theta\left(-\sin k_1-\sin k_2+\sin(k_1-k_2)\right)
\,,
\end{split}   
\end{equation}
where
$\sigma_{\mu}$ ($\mu = 0,1,2,3$) represents the $2 \times 2$ identity
and 
three Pauli matrices, and $\bs{k} = \frac{k_1}{2\pi}\bs{g}_1+\frac{k_2}{2\pi}\bs{g}_2$ ($\bs{g}_1,\bs{g}_2$ are the reciprocal vectors). 
The lowest energy band 
$\varepsilon(\bs{k}) = h_{0}(\bs{k}) - \left({\sum^{}_{i}h^{2}_{i}(\bs{k})}\right)^{1/2}$
has parabolic dispersion $\varepsilon_{\bs{p}} \approx \varepsilon_{0} + \bs{p}^2/(2m^{*})$ near its minimum with effective mass
\begin{equation}
\label{eq: effective mass}
m^* = \frac{m^{*}_{0}}{1 + 6\,\tau_1\,\cos{\theta}}
\,,
\end{equation}
where $m^{*}_{0} = 2\hbar^2/(3|t_0|)$.

\begin{figure*}[!htb]
    \centering
    \includegraphics[width = 18 cm]{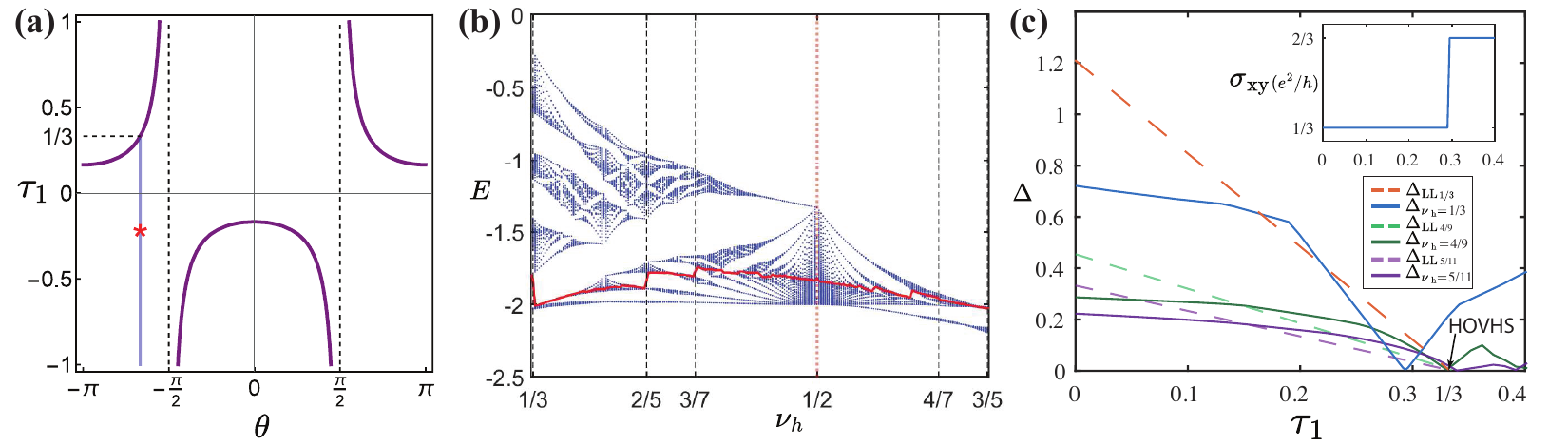}
    \caption{QPT correlated with the HOVHS at $\nu_{\,\text{h}}=1/2$. (a) Curve shows the boundaries where the inverse of effective mass at $\nu_{\,\text{h}}=1/2$ is 0, as a function of $\theta$ and $\tau_1=t_1/|t_0|$, the star symbol indicates the set of $\theta,\tau_1$ used in Fig.\,\ref{Fig2}(a); (b) The Hofstadter spectrum at $t_0=-1,\tau_1=1/3$ and $\theta=-2\pi/3$, zoomed in to show that the composite fermion gaps are nearly closed; 
    {(c) The composite fermion gap versus $\tau_1$: 
    solid and dashed curves are, respectively, the CF gaps and the effective mass prediction \eqref{eq: Delta LL}, at three different fillings $\nu_{\,\text{h}}=1/3,4/9,5/11$. The gap closing points converge to $\tau_1=1/3$ where the HOVHS emerges.}
    The inset plots the transition of the Hall conductance $\sigma_{\text{xy}}$ for the $\nu_{h} = 1/3$ state.
    }
    \label{Fig3}
\end{figure*}

{ 
An intuitive account of the Jain-FCI states then follows. 
From Eq.\,\eqref{eq:flux density relation}, a small change 
$\delta\nu_{\,\text{h}} = \nu_{\,\text{h}} - 1/2~ \ll \mathcal{O}(1)$
away from half-filling 
is connected to a perturbation in the Chern-Simons field $\delta B$, 
\begin{equation}
\delta \nu_{\,\text{h}} = \frac{1}{2}\delta \Phi/\Phi_0 = \frac{1}{2}\delta B\cdot\mathcal{A}_{u.c.}/\Phi_0
\end{equation}

where $\mathcal{A}_{u.c.}=\frac{3\sqrt{3}}{2}$ is the unit cell area. 
This extra field $\delta B$, 
in turn, gives rise to a Landau fan 
with characteristic energy splitting
\begin{equation}
\label{eq: Delta LL}
    \Delta_{\textbf{LL}}  = \hbar \frac{e \, \delta B}{m^*} = \frac{2\Phi_0}{\mathcal{A}_{u.c.}} \frac{\hbar e}{m^*}\left|\nu_{\,\text{h}}-1/2\right|
    \,.
\end{equation}
}

This Landau fan structure is clearly seen in Fig.\,\ref{Fig2}(a), and Eq.\,\eqref{eq: Delta LL} provides the correct energy scales as $\nu_{\,\text{h}} \rightarrow \frac{1}{2}$ \cite{Supplemental}. 
Furthermore, according to Eq.\,\eqref{eq: Delta LL}, 
$\Delta_{\textbf{LL}} \rightarrow 0$  as $m^{*} \rightarrow \infty$, when the band curvature vanishes near the minimum, signaling the onset of a higher order Van Hove singularity (HOVHS)\cite{shtyk2017electrons,yuan2019magic},
which forms along the $\tau_1\cos{\theta} = -1/6$ curves shown in Fig.\,\ref{Fig3}(a), where the denominator of Eq.\,\eqref{eq: effective mass} vanishes. 
Except for $\theta = \pm \pi/2$, a HOVHS occurs for a certain finite ratio of first and second neighbor hopings; for tMoTe$_2$ with $\theta = -2\pi/3$, this occurs for $\tau_1 = 1/3$, shown in the solid vertical line in Fig.\,\ref{Fig3}(a). 

{
Fig.\,\ref{Fig3}(b) displays the Hofstadter spectrum at the HOVHS for 
$\tau_{\textrm{\tiny{HOVHS}}}=1/3$, where a significant reduction in gaps {in the range $\nu_{\,\text{h}}=1/3,2/5\ldots, 3/5$} is observed, associated with the collapsing of the scale $\Delta_{\textbf{LL}}$.

This feature captures the influence of lattice effects on the structure of 
CFs in Chern bands, in stark contrast with Landau levels \cite{halperin1993theory,son2015composite}. While the role of HOVHS in promoting competing electronic orders in Chern bands has been recently emphasized\cite{castro2023emergence,aksoy2023single,pullasseri2024chern,wu2023pair}, to our knowledge, the connection between HOVHS and CF
states has not received earlier consideration, and is one of the central results of this work.
}

{
Remarkably, the proximity to a HOVHS provides a scenario to explore QPTs for a \textit{group} of composite fermion bands
induced by closing of their topological gaps due to lattice effects. To test this scenario, we plot in Fig.\,\ref{Fig3}(c) the 
CF gap of the $\nu_{\,\text{h}}=1/3,4/9,5/11$ states as a function of $\tau_1$ for $\theta=-2\pi/3$. 
Near
$\tau_{\textrm{\tiny{HOVHS}}}=1/3$, we observe these gaps
following a trend similar to that predicted by Eq.\,\eqref{eq: Delta LL}, with the gap closing 
approaching $\tau_{\textrm{\tiny{HOVHS}}}$ as $|\delta\nu_h| \rightarrow 0$, as seen at $\nu_h = 4/9\approx 0.44$ and $\nu_h =  5/11\approx 0.45$.
The deviation at $\nu_h = 1/3$ represents a correction to the asymptotic expression Eq.~(\ref{eq: Delta LL}) due to $\delta\nu_h \sim \mathcal{O}(1)$.
For the 1/3 state, the inset shows that this gap closing marks a QPT between a Jain-FCI (C = 1) and a fractal-FCI (C = -2) state with Hall conductances $e^2/3h$ and $2e^2/3h$, respectively. 
{We have also observed similar QPTs
for other FCI states such as $\nu_{\,\text{h}}=2/5, 3/7, 6/13$ \cite{Supplemental}, confirming the generality of the quantum criticality in proximity to a HOVHS.}
}

\begin{figure}[!htb]
    \centering
    \includegraphics[width = 9 cm]{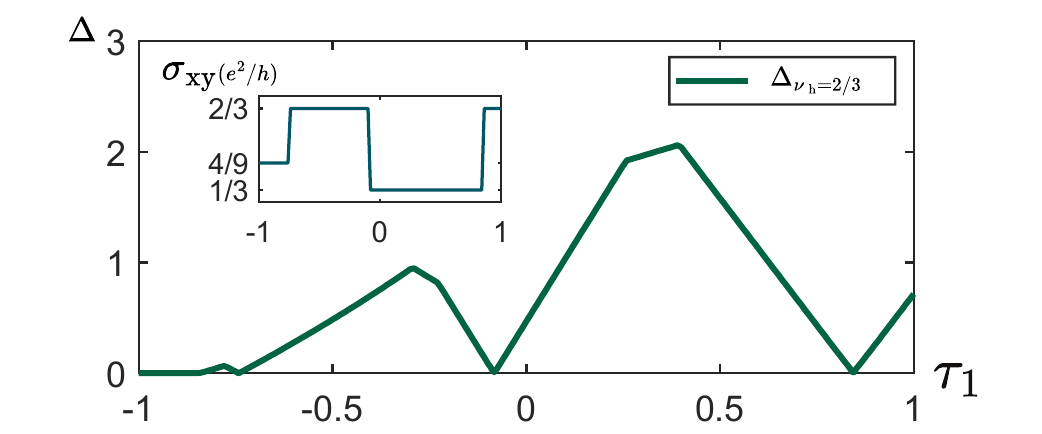}
    \caption{Composite fermion gap $\Delta$ as a function of $\tau_1$ at $\nu_{\,\text{h}}=2/3$ showing phase transitions at $\Delta=0$. $\theta=-2\pi/3, t_0=-1$. The inset plots the phase transition of $\sigma_{\text{xy}}$. The domain of $\sigma_{\text{xy}}=2/3$ ($\sigma_{\text{xy}}=1/3$) corresponds to $-0.74<\tau_1<-0.08$ ($-0.08<\tau_1<0.84$).}
    \label{Fig4}
\end{figure}

We note that a fractal Hofstadter spectrum can also support QPTs outside the HOVHS mechanisms due to individual band touchings.
In fact, an important case occurs at $\nu_{\,\text{h}}=2/3$, which does not follow the Landau fan emanating from the compressible state at half-filling. 
In Fig.\,\ref{Fig4}, we trace the CF
gap and Hall conductance at $\nu_{\,\text{h}}=2/3$ in a wide range of $\tau_1$. QPTs are observed at multiple ratios 
$\tau_1 = t_1/|t_0|$.
Notably, for $ -0.74 < \tau_1 < -0.08$, the incompressible state is a Jain-FCI with Hall conductance $2{e^2}/3h$ topologically consistent with experiments.\cite{cai2023signatures,zeng2023thermodynamic, park2023observation, xu2023observation} 
Furthermore, our analysis predicts that interesting behavior may emerge at $\tau_1 \approx -0.08$ due to competition between distinct FCI states, opening new opportunities for experimental and numerical investigations.

\textit{Discussion--} In summary, employing a composite fermion theory, we have characterized the topological properties of fractional Chern insulators 
in tMoTe$_2$. 
In addition to 
the experimentally observed FCI states at $2/3$ and $3/5$ fillings, our approach reveals several other candidate FCI states, 
suggesting new 
pathways to realize topologically ordered phases sans external magnetic fields.
We highlight the influence of lattice effects on the stability of fractal CF states and their role in inducing quantum phase transitions through higher-order Van Hove singularities.
The development of an effective Haldane-type model for the low-energy bands of tMoTe$_2$ opens the door for future numerical exploration of FCIs via large-scale density matrix renormalization group simulations\cite{Cincio2013topological,Grushin2015stability,Zeng2018fractional,andrews2021stability}.
Expanding this approach to characterize FCIs in multilayer graphene heterostructures \cite{lu2024fractional} is a promising avenue. Furthermore, a time-reversal symmetric generalization of the composite fermion approach can shed light on the experimental evidence\cite{Kang2024evidence} for the fractional quantum spin Hall effect \cite{bernevig2006quantum,levin2009fractional,santos2011time,neupert2011fractional} in moir\'e tMoTe$_2$. We leave these open questions for future investigation.
\vspace{4pt}

\textit{Acknowledgments--} We are grateful to Andrei Bernevig, Jainendra Jain, and Nicolas Regnault for stimulating discussions. This research was supported by the U.S. Department of Energy, Office of Science, Basic Energy Sciences, under Award DE-SC0023327. 

\bibliography{bibliography}

\clearpage
\widetext

\begin{center}
\textbf{\Large Supplemental Materials}
\vskip 1em%
{\normalsize
      \lineskip .5em%
      \begin{tabular}[t]{c}%
      Tianhong\,\,Lu$^{1}$\,\,and\,\,Luiz\,\,H.\,\,Santos$^{1}$
        \vspace{0.4 em}
        \\{\small
        $^{1}$\textit{Department  of  Physics,  Emory  University,  400 Dowman Drive, Atlanta,  GA  30322,  USA}}
      \end{tabular}\par}%
    {\small (Dated: Aug 16, 2024)}%
\end{center}
\par
\vskip 1.5em
\renewcommand{\theequation}{S\arabic{equation}}
\setcounter{equation}{0}
\renewcommand{\thefigure}{S\arabic{figure}}
\renewcommand{\thetable}{S\arabic{table}}
\setcounter{figure}{0}
\setcounter{table}{0}
\setcounter{secnumdepth}{4}

Sec.\,\ref{Sec1} describes the Haldane Hamiltonian for the two top bands of tMoTe$_2$. In Sec.\,\ref{Sec2}, we derive the effective Hamiltonian coupled to the Chern-Simons gauge field. Sec.\,\ref{Sec3} discussed the Chern-Simons theory of flux attachment and its relation to composite fermions. 
In Sec.\,\ref{Sec4}, we provide details of the analytical calculation of the effective mass at half-filling ($\nu_{\,\text{h}}=1/2$), and perform linear regression on the Hofstadter spectrum close to half-filling. Furthermore, we provide three more examples of Fig.\,\ref{Fig3}(c) at $\nu_{\,\text{h}}=2/5,3/7,6/13$, as well as generalize the HOVHS to the $\tau_2\neq0$ scenario.

\section{Effective Hamiltonian in the real space}
\label{Sec1}

\begin{figure}[!htb]
    \centering
    \includegraphics[width=0.7\linewidth]{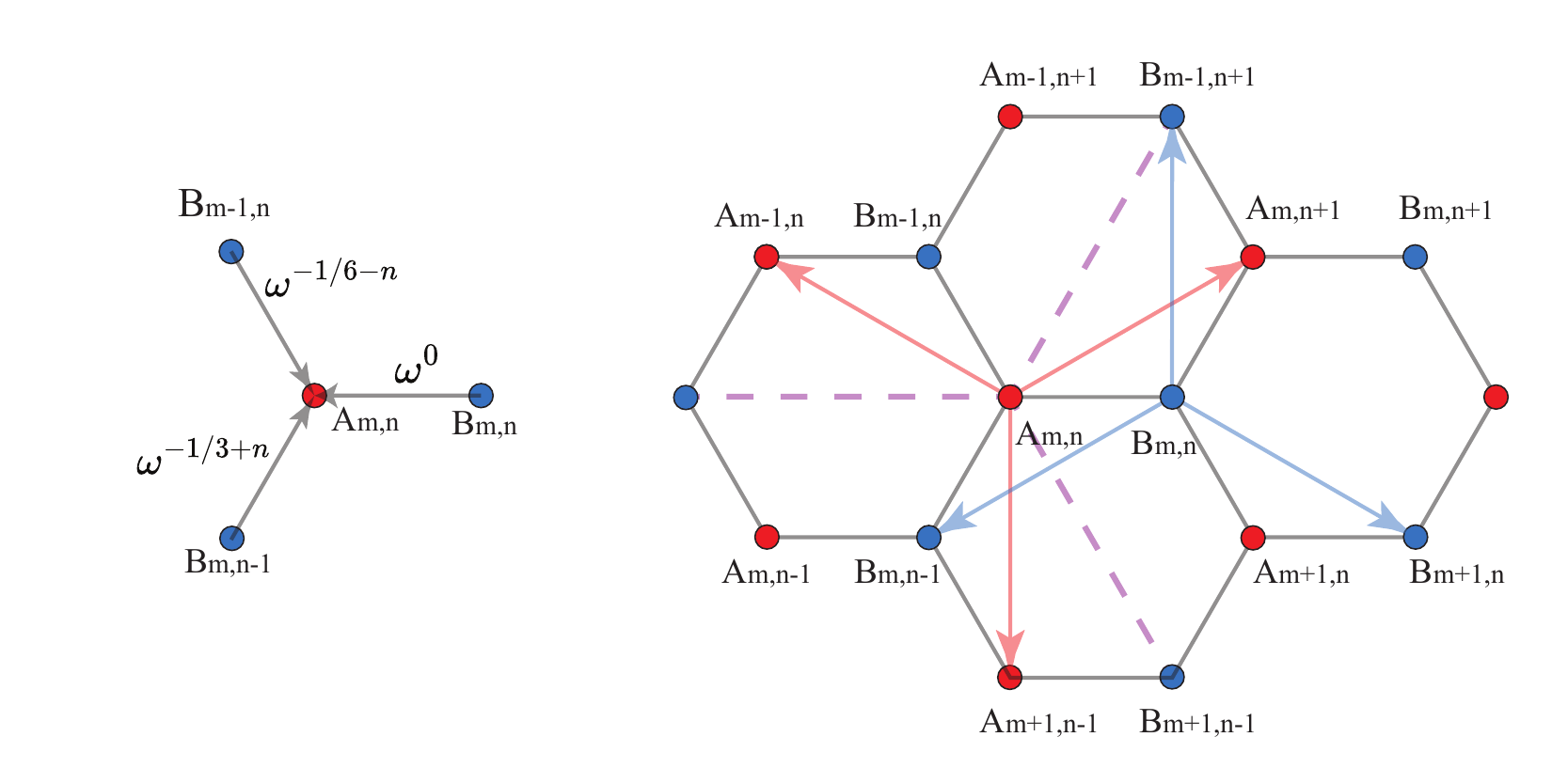}
    \caption{Left: arrows indicate direction of hopping (i.e. towards site $A_{m,n}$) and the Peierls phase accumulated. Right: arrows indicate direction of NNN hopping amplitude $t_1 e^{i\theta}$.
     Dashed line denotes the third-neighbor hopping. 
    The accumulated Peierls phase in along the NNN hopping can be directly obtained from the phases on the left and the fact that the Aharonov-Bohm phase experience by a charged particle is $e^{\pm i 2\pi \phi_{\text{triangle}}/\phi_0}$ where $\pm $ indicates CCW/CW dirctions and $\phi_{\text{triangle}} = \phi_0/6$ is the flux in a triangle that has a blue arrow as one of the edges.}
    \label{Supp: fig: Honeycomb lattice}
\end{figure}

We chose nearest neighbor vectors to be $\bs{\xi}_1 = a(1,0)$, $\bs{\xi}_2 = a(-1/2,+\sqrt{3}/2)$, $\bs{\xi}_3 = a(-1/2,-\sqrt{3}/2)$. Henceforth, $a=1$. The primitive vectors of the Bravais lattice are $\bs{a}_1 = \bs{\xi}_1 - \bs{\xi}_2 = (3/2, -\sqrt{3}/2)$ and $\bs{a}_2 = \bs{\xi}_1 - \bs{\xi}_3 = (3/2, \sqrt{3}/2)$. The lattice vectors are thus $\bs{r} = m\,\bs{a}_1 + n\,\bs{a}_2$ with $m,n \in \mathbb{Z}$. We arrange the honeycomb lattice such that $A_{m,n} = \psi_{\bs{r} = m\,\bs{a}_1 + n\,\bs{a}_2}$ and $B_{m,n} = \psi_{\bs{r} + \bs{t}_1 = m\,\bs{a}_1 + n\,\bs{a}_2 + \bs{t}_1}$., as shown in the Fig.\,\ref{Supp: fig: Honeycomb lattice}.

\subsection{The origin of the phase factor}
The low energy description of tMoTe$_2$ can be modeled via a two-layer spin-polarized continuum model\cite{wu2019topological}, take spin  $\sigma=1$ for example,
\begin{equation}
    \mathcal{H}_{\uparrow}=\left(\begin{array}{cc}-\frac{\hbar^2\left(\boldsymbol{k}-\boldsymbol{\kappa}_{+}\right)^2}{2 m^*}+\Delta_{\mathfrak{b}}(\boldsymbol{r}) & \Delta_T(\boldsymbol{r}) \\ \Delta_T^{\dagger}(\boldsymbol{r}) & -\frac{\hbar^2\left(\boldsymbol{k}-\boldsymbol{\kappa}_{-}\right)^2}{2 m^*}+\Delta_{\mathfrak{t}}(\boldsymbol{r})\end{array}\right)
\end{equation}
where $\boldsymbol{\kappa}_{\pm} = (-\sqrt{3},\pm1)\cdot2\pi/(3a_0)\theta$
is the relative momentum shift caused by the twisting ($\theta$) of the two layers, and $a_0$ is the lattice constant of mono-layer MoTe$_2$.

The momentum shift produces a phase factor 
$\exp(-i\boldsymbol{\kappa_\pm}\cdot\boldsymbol{r}) c_{\ell=\pm, \sigma=1, \boldsymbol{r}}$
in the tight binding model (\ref{eq: full TB model}) when hopping within the same sublattices (same layers). Take counterclockwise AA hopping ($t_1$) for example, $\boldsymbol{\delta_1}\in\{\vec{a_2},-\vec{a_1},\vec{a_1}-\vec{a_2}\}$ where $|\delta_1| = a_0/\theta$
connect two A sites. The full hopping term is $t_1\exp(i\boldsymbol{\kappa_+}\cdot\boldsymbol{\delta_1}) c_{+, \sigma=1, \boldsymbol{r+\delta_1}}^{\dagger}c_{+, \sigma=1, \boldsymbol{r}} $, where
\begin{equation}
    \begin{split}
    &\boldsymbol{\kappa_+}\cdot \vec{a_2}
    =(-\sqrt{3},1)\cdot2\pi/(3a_0)\theta
    \cdot
    (\sqrt{3}/2,1/2)a_0/\theta = -2\pi/3\\
    &\boldsymbol{\kappa_+}\cdot (-\vec{a_1})
    =(-\sqrt{3},1)\cdot2\pi/(3a_0)\theta
    \cdot
    (-\sqrt{3}/2,1/2)a_0/\theta = 4\pi/3=-2\pi/3\\
    &\boldsymbol{\kappa_+}\cdot (\vec{a_1}-\vec{a_2})
    =(-\sqrt{3},1)\cdot2\pi/(3a_0)\theta
    \cdot
    (0,-1)a_0/\theta = -2\pi/3\\
    \end{split}
\end{equation}
which give rise to the phase $\theta_{\sigma=1}^{\,e}=-2\pi/3$. The other spin $\sigma=-1$ can be mapped to by the time reversal transformation, which can be summarized as $\theta_{\sigma}^{\,e}=-2\pi\sigma/3$.

\subsection{Two-band Haldane model}

The model contains a NN hopping of strength $t_0$ and a complex hopping $t_1\,e^{i\theta}$ as indicated in Fig.\,\ref{Supp: fig: Honeycomb lattice} (right),
\begin{equation}
H = H_0 + H_1 \equiv H_0 + H_{AA} + H_{BB}
\,,
\end{equation}
where
\begin{equation}
\label{eq: NN H 01}
H_{0} = t_{0}\,
\sum_{m,n} 
\Big(
A^{\dagger}_{m,n}\,B_{m,n} + 
A^{\dagger}_{m,n}\,B_{m-1,n}
+
A^{\dagger}_{m,n}\,B_{m,n-1}
\Big)
+
\textrm{H.c.}
\,,
\end{equation}
\begin{equation}
H_{AA} = t_1\,e^{i\theta}\,\sum_{m,n} 
\Big(A^{\dagger}_{m,n+1} A^{}_{m,n}
+
A^{\dagger}_{m-1,n} A^{}_{m,n}
+
A^{\dagger}_{m+1,n-1} A^{}_{m,n}
\Big)
+
\textrm{H.c.}
\,
\end{equation}
and
\begin{equation}
H_{BB} = t_1\,e^{i\theta}\,\sum_{m,n} 
\Big(B^{\dagger}_{m-1,n+1} B^{}_{m,n}
+
B^{\dagger}_{m,n-1} B^{}_{m,n}
+
B^{\dagger}_{m+1,n} B^{}_{m,n}
\Big)
+
\textrm{H.c.}
\,.
\end{equation}

Defining 
the Fourier transform of the fermionic operators as 
\begin{equation}
A_{\bs{r}} \equiv A_{m,n}
=
\frac{1}{\sqrt{N}}\sum_{\bs{k} \in BZ}\,e^{i \bs{k} \cdot \bs{r}}A_{\bs{k}}
\,,\quad
B_{\bs{r}}
\equiv
B_{m,r}
=
\frac{1}{\sqrt{N}}\sum_{\bs{k} \in BZ}\,e^{i \bs{k} \cdot \bs{r}}B_{\bs{k}}
\,,
\end{equation}
leads to 
\begin{equation}
H_0 = \sum_{\bs{k}} 
A^{\dagger}_{\bs{k}}B_{\bs{k}} \,t_{0}(1 + e^{-i k_1} + e^{-i k_2}) + \textrm{H.c.}
\,,
\end{equation}
\begin{equation}
H_{AA} = \sum_{\bs{k}} 
A^{\dagger}_{\bs{k}} A_{\bs{k}} \,
t_{1}e^{i\theta}(e^{-i k_2} + e^{i k_1} + e^{i(-k_1 + k_2)})  + \textrm{H.c.}
\end{equation}
\begin{equation}
H_{BB} = \sum_{\bs{k}} 
B^{\dagger}_{\bs{k}} B_{\bs{k}} \,t_{1}e^{i\theta}(e^{i k_2} + e^{-i k_1} + e^{i(k_1 - k_2)})  + \textrm{H.c.}
\end{equation}

All in all, the model is written as
\begin{equation}
H = 
\sum_{\bs{k}}
\begin{pmatrix}
A^{\dagger}_{\bs{k}} & B^{\dagger}_{\bs{k}} \end{pmatrix}
\,
\begin{pmatrix}
t_{1}e^{i\theta}(e^{-i k_2} + e^{i k_1} + e^{i(-k_1 + k_2)})  + \textrm{c.c.}
&
t_{0}(1 + e^{-i k_1} + e^{-i k_2})
\\
t_{0}(1 + e^{i k_1} + e^{i k_2})
&
t_{1}e^{i\theta}(e^{i k_2} + e^{-i k_1} + e^{i(k_1 - k_2)})  + \textrm{c.c.}
\end{pmatrix}
\begin{pmatrix}
A_{\bs{k}} \\ B_{\bs{k}} \end{pmatrix}
\end{equation}

\section{Hofstadter bands}
\label{Sec2}
Perpendicular to the plane of the lattice we consider is a uniform Chern-Simons field $\bs{B}  = \bs{\nabla} \times \bs{a}$ and we adopt the gauge $\bs{a} = B(x + \sqrt{3}y)\vec{e}_y$. The area of the unit cell is $A_{u.c.} = 3\sqrt{3}/2$ which gives a flux $\phi = B\,3\sqrt{3}/2$. With $B$ in the positive direction, this gives a positive counter-clockwise phase $e^{i (q/\hbar) \oint d \bs{l} \cdot \bs{a}} = e^{i (q/\hbar) \phi} = e^{i 2\pi \phi/\phi_0}$, where $\phi_0 = q/h$. Henceforth, we define $\omega = e^{i 2\pi \phi/\phi_0}$ and we will be considering the case where $\phi/\phi_0 = p/q$ is a rational number.

We describe the effect of the Chern-Simons gauge field via a Peierls substitution according to 
\begin{equation}
t_{\bs{r},\bs{r}'}\,\psi^{\dagger}_{\bs{r}}\psi_{\bs{r}'} \rightarrow
t_{\bs{r},\bs{r}'}\,
e^{i (2\pi/\phi_0) \int^{\bs{r}}_{\bs{r}'} d \bs{l} \cdot \bs{a}}
\,
\psi^{\dagger}_{\bs{r}}\psi_{\bs{r}'} 
\,.
\end{equation}

\subsection{Nearest neighbor hopping}

Fig.\,\ref{Supp: fig: Honeycomb lattice} (left) shows the Peiers phases accumulated a charged particle that hops into the $A_{m,n}$ site from its nearest neighbor B sites. This leads to the tight-binding Hamiltonian
\begin{equation}
\label{eq: NN H 02}
H_{0} = t_{0}\,
\sum_{m,n} 
\Big(
A^{\dagger}_{m,n}\,B_{m,n} + 
\omega^{-1/6-n}\,
A^{\dagger}_{m,n}\,B_{m-1,n}
+
\omega^{-1/3+n}\,
A^{\dagger}_{m,n}\,B_{m,n-1}
\Big)
+
\textrm{H.c.}
\end{equation}

We see that the Hamiltonian has an effective translation symmetry $(m,n) \rightarrow (m+1,n+q)$, which embodies the  magnetic translation symmetry of the Hamiltonian in the presence of an external magnetic field. Naturally, we extend the unit cell along the $\bs{a}_2$ direction such that the system has a magnetic unit cell with $2\times q$ sites and the primitive lattice vectors are $\bs{a}^{'}_1 = \bs{a}^{}_1$ and $\bs{a}^{'}_2 = q\,\bs{a}^{}_2$. Accordingly, we define reciprocal lattice vectors of the magnetic BZ as $\bs{g}^{'}_1$ and $\bs{g}^{'}_2$  via $\bs{a}^{'}_i \cdot \bs{g}^{'}_j = 2\pi \delta_{ij}$, and parameterized momentum via $\bs{k} = k_1 \bs{g}^{'}_1 + k_2 \bs{g}^{'}_2$.

Letting $n = q\,r + s$ with $r \in \mathbb{Z}$ and $s = 0, ..., q-1$, we introduce subalttice operators as
\begin{equation}
A_{m, n} = A_{m, q\,r + s} \equiv A^{(s)}_{m,r} 
\equiv A^{(s)}_{\bs{R}} 
\,,
\quad
B_{m, n} = B_{m, q\,r + s} \equiv B^{(s)}_{m,r} \equiv B^{(s)}_{\bs{R}}  
\,,
\end{equation}
where $\bs{R} = m\bs{a}^{'}_{1} + r\bs{a}^{'}_2$ is the position of the magnetic unit cell, and their Fourier transforms
\begin{equation}
A^{(s)}_{m,r}
=
\frac{1}{\sqrt{N}}\sum_{\bs{k} \in MBZ}\,e^{i \bs{k} \cdot \bs{R}}A^{(s)}_{\bs{k}}
\,,\quad
B^{(s)}_{m,r}
=
\frac{1}{\sqrt{N}}\sum_{\bs{k} \in MBZ}\,e^{i \bs{k} \cdot \bs{R}}B^{(s)}_{\bs{k}}
\end{equation}
and express each of the terms  of the Hamiltonian \eqref{eq: NN H 02} as
\begin{equation}
X_1 = \sum_{m,n} 
A^{\dagger}_{m,n}\,B_{m,n}
=
\sum_{m,r}\sum^{q-1}_{s=0}\,
(A^{s}_{m,r})^{\dagger}\,B^{s}_{m,r}
=
\sum_{\bs{k}}\sum^{q-1}_{s=0}\,
(A^{s}_{\bs{k}})^{\dagger}\,B^{s}_{\bs{k}}
\,,
\end{equation}
\begin{equation}
X_2 = \sum_{m,n}
\omega^{-1/6-n}
A^{\dagger}_{m,n}\,B_{m-1,n}
=
\sum_{m,r}\sum^{q-1}_{s=0}\,
\omega^{-1/6-s}
(A^{s}_{m,r})^{\dagger}\,B^{s}_{m-1,r}
=
\sum_{\bs{k}}\sum^{q-1}_{s=0}\,
\omega^{-1/6-s}
e^{-ik_1}
(A^{s}_{\bs{k}})^{\dagger}\,B^{s}_{\bs{k}}
\,,
\end{equation}
and
\begin{equation}
\begin{split}
X_3
&\,=
\sum_{m,n}
\omega^{-1/3+n}\,
A^{\dagger}_{m,n}\,B_{m,n-1}
=
\sum_{m,r}\sum^{q-1}_{q=0}
\omega^{-1/3+s}\,
A^{\dagger}_{m,qr+s}\,B_{m,qr+s-1}
\\
&\,=
\sum_{m,r}
\Big[
\omega^{-1/3}\,
A^{\dagger}_{m,qr}\,B_{m,qr-1}
+
\sum^{q-1}_{q=1}
\omega^{-1/3+s}\,
A^{\dagger}_{m,qr+s}\,B_{m,qr+s-1}
\Big]
\\
&\,=
\sum_{m,r}
\Big[
\omega^{-1/3}\,
(A^{(0)}_{m,r})^{\dagger}\,B^{(q-1)}_{m,r-1}
+
\sum^{q-1}_{q=1}
\omega^{-1/3+s}\,
(A^{(s)}_{m,r})^{\dagger}\,B^{(s-1)}_{m,r}
\Big]
\\
&\,=
\sum_{\bs{k}}
\Big[
\omega^{-1/3}\,
e^{-i k_2}
(A^{(0)}_{\bs{k}})^{\dagger}\,B^{(q-1)}_{\bs{k}}
+
\sum^{q-1}_{q=1}
\omega^{-1/3+s}\,
(A^{(s)}_{\bs{k}})^{\dagger}\,B^{(s-1)}_{\bs{k}}
\Big]
\end{split}
\end{equation}

Then the nearest neighbor Hamiltonian $H_0 = t_0\,(X_1+X_2+X_3) + \textrm{H.c.}    
$ then reads
\begin{equation}
\begin{split}
H_0
&\,= 
t_0
\sum_{\bs{k}}\sum^{q-1}_{s=0}\,
(1 + \omega^{-1/6-s}
e^{-ik_1})
(A^{s}_{\bs{k}})^{\dagger}\,B^{s}_{\bs{k}}  + \textrm{H.c.}
\\
&\,
t_0
\sum_{\bs{k}}
\Big[
\omega^{-1/3}\,
e^{-i k_2}
(A^{(0)}_{\bs{k}})^{\dagger}\,B^{(q-1)}_{\bs{k}}
+
\sum^{q-1}_{q=1}
\omega^{-1/3+s}\,
(A^{(s)}_{\bs{k}})^{\dagger}\,B^{(s-1)}_{\bs{k}}
\Big]
+ \textrm{H.c.}
\end{split}
\end{equation}

\subsection{Next-nearest neighbor hopping}

The next-nearest neighbor Hamiltonian 
\begin{equation}
H_1 = H_{AA} + H_{BB}    
\end{equation}
corresponds to hopping between equal sublattices.

The AA hopping - with the arrows in Fig.\,\ref{Supp: fig: Honeycomb lattice} (right) indicates the direction with topological hopping $t_1 e^{i\theta}$ -  is given by
\begin{equation}
H_{AA} = t_1\,e^{i\theta}\,\sum_{m,n} 
\Big(\omega^{n+1/2}\,A^{\dagger}_{m,n+1} A^{}_{m,n}
+
\omega^{n}\,A^{\dagger}_{m-1,n} A^{}_{m,n}
+
\omega^{1-2n}
A^{\dagger}_{m+1,n-1} A^{}_{m,n}
\Big)
+
\textrm{H.c.}
\end{equation}
Fourier transforming,
\begin{equation}
\begin{split}
H_{AA}
&\,=
t_1 e^{i\theta}
\sum_{\bs{k}}
\Big[
\sum^{q-2}_{s=0}
\omega^{s+1/2} A^{s+1\dagger}_{\bs{k}} A^{s}_{\bs{k}}
+
\omega^{(q-1)+1/2} e^{-i k_2}
A^{0\dagger}_{\bs{k}} A^{q-1}_{\bs{k}}
\Big]
\\
&\,+
t_1 e^{i\theta}
\sum_{\bs{k}}\sum^{q-1}_{s=0}
\omega^{s} e^{i k_1}
A^{s\dagger}_{\bs{k}} A^{s}_{\bs{k}}
\\
&\,+
t_1 e^{i\theta}
\sum_{\bs{k}} 
\Big[
\omega^{1} e^{-i k_1 + i k_2} 
A^{q-1\dagger}_{\bs{k}} A^{0}_{\bs{k}} 
+ 
\sum^{q-1}_{s=1}\omega^{1-2s} e^{-i k_1} A^{s-1 \dagger}_{\bs{k}} A^{s}_{\bs{k}} 
\Big]
\\
&\,+ \textrm{H.c.}
\end{split}    
\end{equation}

The BB hopping - with the arrows in Fig.\,\ref{Supp: fig: Honeycomb lattice} (right) indicates the direction with topological hopping $t_1 e^{i\theta}$ - is given by
\begin{equation}
H_{BB} = t_1\,e^{i\theta}\,\sum_{m,n} 
\Big(\omega^{2n+5/3}\,B^{\dagger}_{m-1,n+1} B^{}_{m,n}
+
\omega^{1/6 - n}\,B^{\dagger}_{m,n-1} B^{}_{m,n}
+
\omega^{-1/3-n}
B^{\dagger}_{m+1,n} B^{}_{m,n}
\Big)
+
\textrm{H.c.}
\end{equation}

By Fourier transform, we get 
\begin{equation}
\begin{split}
H_{BB}
&\,=
t_1 e^{i\theta}
\sum_{\bs{k}}
\Big[
\sum^{q-2}_{s=0} \omega^{5/3 + 2s} e^{i k_1} B^{s+1\dagger}_{\bs{k}} B^{s}_{\bs{k}}
+
\omega^{5/3+2(q-1)} e^{i k_1-i k_2} B^{0\dagger}_{\bs{k}} B^{q-1}_{\bs{k}}
\Big]
\\
&\,+
t_1 e^{i\theta}
\sum_{\bs{k}}
\Big[
\sum^{q-1}_{s=1} \omega^{1/6 - s} B^{s-1\dagger}_{\bs{k}} B^{s}_{\bs{k}}
+
\omega^{1/6} e^{i k_2} B^{q-1\dagger}_{\bs{k}} B^{0}_{\bs{k}}
\Big]
\\
&\,+
t_1 e^{i\theta}
\sum_{\bs{k}}
\sum^{q-1}_{s=0} \omega^{-1/3 - s} e^{-i k_1}
B^{s\dagger}_{\bs{k}} B^{s}_{\bs{k}}
\\
&\, + \textrm{H.c.}
\end{split}
\end{equation}

\subsection{Third-nearest neighbor hopping}
The third-nearest neighbor Hamiltonian
\begin{equation}
    H_2 = t_2 
    \sum_{m, n}\left(
    \omega^{2n+4/3}B_{m-1, n+1}^{\dagger} A_{m, n}+
    \omega^{2/3-2n}B_{m+1, n-1}^{\dagger} A_{m, n}+
    \omega^{0}B_{m-1, n-1}^{\dagger} A_{m, n}\right)+\text { H.c. }
\end{equation}
corresponds to hopping between sublattices across the hexagon as denoted in Fig.\,\ref{Supp: fig: Honeycomb lattice}.

Fourier transforming,
\begin{equation}
    H_2 = t_2 (X_1+X_2+X_3)+\text { H.c. }
    \label{H2 eq}
\end{equation}
where
\begin{equation}
    \begin{split}
        X_1 &= \sum_{m, n}
    \omega^{2n+4/3}B_{m-1, n+1}^{\dagger} A_{m, n} 
    = \sum_{m, r}\sum_{s=0}^{q-1}
    \omega^{2s+4/3}B_{m-1,qr+s+1}^{\dagger}A_{m,qr+s}\\
    &=\sum_{m, r}(\omega^{2(q-1)+4/3}B_{m-1,qr+q}^{\dagger}A_{m,qr+q-1}
    +
    \sum_{s=0}^{q-2}
    \omega^{2s+4/3}B_{m-1,qr+s+1}^{\dagger}A_{m,qr+s})\\
    &= \sum_{\boldsymbol{k}}
    \sum_{s=0}^{q-2}
    \omega^{2s+4/3}
e^{ik_1}B_{\boldsymbol{k}}^{(s+1)\dagger}A_{\boldsymbol{k}}^{(s)}+\omega^{-2/3}e^{ik_1-ik_2}B_{\boldsymbol{k}}^{(0)\dagger}A_{\boldsymbol{k}}^{(q-1)}
    \end{split}
\end{equation}
\begin{equation}
    \begin{split}
        X_2 &= \sum_{m, n}
    \omega^{2/3-2n}B_{m+1, n-1}^{\dagger} A_{m, n}=
    \sum_{m, r}\sum_{s=0}^{q-1}
    \omega^{2/3-2s}B_{m+1, qr+s-1}^{\dagger} A_{m, qr+s}\\
    &=\sum_{m, r}(
    \sum_{s=1}^{q-1}
    \omega^{2/3-2s}B_{m+1, qr+s-1}^{\dagger} A_{m, qr+s}+
    \omega^{2/3}B_{m+1, qr-1}^{\dagger} A_{m, qr+0}
    )\\
    &=\sum_{\boldsymbol{k}}(
    \sum_{s=1}^{q-1}
    e^{-ik_1}
    \omega^{2/3-2s}B^{(s-1)\dagger}_{\boldsymbol{k}} A^{(s)}_{\boldsymbol{k}}+
    \omega^{2/3}
    e^{-ik_1+ik_2}
    B^{(q-1)\dagger}_{\boldsymbol{k}} A^{(0)}_{\boldsymbol{k}}
    )
    \end{split}
\end{equation}
\begin{equation}
    \begin{split}
        X_3 &= \sum_{m, n}
    B_{m-1, n-1}^{\dagger} A_{m, n}=\sum_{m, r}
    \sum_{s=0}^{q-1}
    B_{m-1, qr+s-1}^{\dagger} A_{m, qr+s}\\
    &=\sum_{m, r}
    \sum_{s=1}^{q-1}
    B_{m-1, qr+s-1}^{\dagger} A_{m, qr+s}+B_{m-1, qr-1}^{\dagger} A_{m, qr+0}\\
    &=\sum_{\boldsymbol{k}}
    \sum_{s=1}^{q-1}
    e^{ik_1}
    B^{(s-1)\dagger}_{\boldsymbol{k}} A^{(s)}_{\boldsymbol{k}}
    +e^{ik_1+ik_2}
    B^{(q-1)\dagger}_{\boldsymbol{k}} A^{(0)}_{\boldsymbol{k}}
    \end{split}
\end{equation}

\section{Chern-Simons theory of flux attachment}
\label{Sec3}
The composite fermion theory characterizes an incompressible state of a partially filled topological band in terms of a band insulator of composite fermions, which are bound states of an electron (or hole) and an even number $2p$ of flux quanta $\phi_0 = h/e = 2\pi$ (in units where $\hbar = e  = 1)$. 

To accomplish the flux attachment, we introduce a Chern-Simons statistical gauge field $a_\mu$ and impose the condition
\begin{equation}
\label{eq:flux attachment}
(2p)\, J^{\mu} = \frac{1}{2\pi} \varepsilon^{\mu\nu\lambda} \partial_{\nu}a_{\lambda}  
\,,
\end{equation}
where $J^{\mu}$ is the number current density. To understand the significance of Eq.\,\ref{eq:flux attachment}, consider the $\mu=0$ equation
\begin{equation}
(2p)\, J^{0} = \frac{1}{2\pi} \varepsilon^{0 i j} \partial_{i}a_{j}  
\,,
\end{equation}
where $J^{0}$ is the number density. Integrating over space and using Stoke's equation
\begin{equation}
(2p)\, \int d^{2} \bs{r}J^{0} = \frac{1}{2\pi} 
\int d^{2} \bs{r}\varepsilon^{0 i j} \partial_{i}a_{j}  
=
\frac{1}{2\pi} \oint \bs{a} \cdot d\bs{\ell}
\,.
\end{equation}
Since the right hand side gives the Chern-Simons flux divided by the flux quantum (recall $\phi_0 = 2\pi$), this equation establishes that each particle is attached to $2p$ flux quanta.

It turns out that Eq.\,\ref{eq:flux attachment} can be obtained as the Euler-Lagrange equation of motion of a \textit{local} field theory as follows. First, because $J^{\mu}$ is a conserved current, that is $\partial_{\mu} J^{\mu} = 0$, it can be expressed in terms of the gradient of a gauge field $b_{\mu}$ as follows,
\begin{equation}
\label{eq: duality}
J^{\mu} \equiv \frac{1}{2\pi} \varepsilon^{\mu\nu\lambda}\partial_{\nu} b_{\lambda}
\,.
\end{equation}
Eq.\,\ref{eq: duality} describes a particle-vortex duality transformation in $(2+1)$ dimensions.

Consider now the action
\begin{equation}
S_1 = \int d^{3}x \mathcal{L}
=
\int d^{3}x 
\Big(
-\frac{2p}{4\pi}
\varepsilon^{\mu\nu\lambda}
b_{\mu}
\partial_{\nu} b_{\lambda}
+
\frac{1}{2\pi}
\varepsilon^{\mu\nu\lambda}a_{\mu}
\partial_{\nu} b_{\lambda}
\Big)
\end{equation}

Then, it is straightforward to verify that 
\begin{equation}
\frac{\delta S_1}{\delta b_{\mu}} = 0
\quad
\Leftrightarrow
\quad
 \frac{2p}{2\pi} \varepsilon^{\mu\nu\lambda}\partial_{\nu} b_{\lambda} = \frac{1}{2\pi} \varepsilon^{\mu\nu\lambda} \partial_{\nu}a_{\lambda}
 \,,
\end{equation}
which is the flux attachment condition Eq.\,\ref{eq:flux attachment}. 

Now, we write general action that contains fermions and the gauge fields
\begin{equation}
S_2 = \int d^{3}x \mathcal{L}
=
\int d^{3}x 
\Big(
\mathcal{L}(\psi^{\dagger},\psi,a_{\mu})
-\frac{2p}{4\pi}
\varepsilon^{\mu\nu\lambda}
b_{\mu}
\partial_{\nu} b_{\lambda}
+
\frac{1}{2\pi}
\varepsilon^{\mu\nu\lambda}a_{\mu}
\partial_{\nu} b_{\lambda}
\Big)
\,
\end{equation}
where $\mathcal{L}(\psi^{\dagger},\psi,a_{\mu})$ is the Lagrangian describing fermions coupled to the statistical gauge field $a_\mu$. When fermions forms a gapped insulator state, integrating out fermions results in a \textit{universal} Chern-Simons response
\begin{equation}
S_3 = \int d^{3}x \mathcal{L}
=
\int d^{3}x 
\Big(
\frac{C}{4\pi}
\varepsilon^{\mu\nu\lambda}a_{\mu}
\partial_{\nu} a_{\lambda}
-\frac{2p}{4\pi}
\varepsilon^{\mu\nu\lambda}
b_{\mu}
\partial_{\nu} b_{\lambda}
+
\frac{1}{2\pi}
\varepsilon^{\mu\nu\lambda}a_{\mu}
\partial_{\nu} b_{\lambda}
\Big)
\,
\end{equation}
where $C$ is the Chern-number of the fermionic insulator, which in this case is the Chern number of the composite fermion state. 

To probe this state's response, we turn on an external probe field $A_\mu$ minimally coupled to the charge current $q\,J^{\mu}$ ($q=-e$ for the electron) by adding the term $-J^{\mu}A_{\mu} = -\,\frac{1}{2\pi}
\varepsilon^{\mu\nu\lambda}A_{\mu}
\partial_{\nu} b_{\lambda}$, which results in the action
\begin{equation}
S_4 = \int d^{3}x \mathcal{L}
=
\int d^{3}x 
\Big(
\frac{C}{4\pi}
\varepsilon^{\mu\nu\lambda}a_{\mu}
\partial_{\nu} a_{\lambda}
-\frac{2p}{4\pi}
\varepsilon^{\mu\nu\lambda}
b_{\mu}
\partial_{\nu} b_{\lambda}
+
\frac{1}{2\pi}
\varepsilon^{\mu\nu\lambda}a_{\mu}
\partial_{\nu} b_{\lambda}
-\frac{1}{2\pi}
\varepsilon^{\mu\nu\lambda}A_{\mu}
\partial_{\nu} b_{\lambda}
\Big)
\,
\end{equation}

We introduce a two-component gauge field
\begin{equation}
a^{I}_{\mu} = 
\begin{pmatrix}
b_\mu
\\
a_{\mu}
\end{pmatrix}
\,,
\end{equation}
and express the action as
\begin{equation}
S_4 = \int d^{3}x \mathcal{L}
=
\int d^{3}x 
\Big(
\frac{1}{4\pi}
K_{I J}
\varepsilon^{\mu\nu\lambda}a^{I}_{\mu}
\partial_{\nu} a^{J}_{\lambda}
-Q_{I}\,\frac{1}{2\pi}
\varepsilon^{\mu\nu\lambda}A_{\mu}
\partial_{\nu} a^{I}_{\lambda}
\Big)
\,
\end{equation}
where
\begin{equation}
K
=
\begin{pmatrix}
-2 p & 1
\\
1 & C
\end{pmatrix}
\,,
\quad
Q = \begin{pmatrix}
1
\\
0
\end{pmatrix}    
\end{equation}
are, respectively, the symmetric K-matrix and the charge vector.

Integrating out $a^{I}$ fields  gives the response action for the  external probe field 
\begin{equation}
S_{eff} = \int d^{3}x \mathcal{L}
=
-
Q^{T}\cdot K^{-1} \cdot Q
\int d^{3}x 
\frac{1}{4\pi}
\varepsilon^{\mu\nu\lambda}A_{\mu}
\partial_{\nu} A_{\lambda}
\,.
\end{equation}
From $J^{\mu} = \frac{\delta S_{eff}}{\delta A_{\mu}}$, one recovers
the Hall conductivity 
\begin{equation}
\sigma_{xy} =
\frac{C}{2p\,C + 1}
\,.
\end{equation}
\section{Composite fermion gaps correlated with emergent HOVHS}
\label{Sec4}
\subsection{Effective mass}

At half-filling ($\nu_{\,\text{h}}=1/2$), $p=3,q=1$, the Hamiltonian is $2\times2$.  
The nearest neighbor Hamiltonian reads
\begin{equation}
\begin{split}
H_0
&\,= 
t_0
\sum_{\bs{k}}\sum^{q-1}_{s=0}\,
(1 + \omega^{-1/6-s}
e^{-ik_1})
(A^{s}_{\bs{k}})^{\dagger}\,B^{s}_{\bs{k}}  + \textrm{H.c.}
\\
&\,
+t_0
\sum_{\bs{k}}
\Big[
\omega^{-1/3}\,
e^{-i k_2}
(A^{(0)}_{\bs{k}})^{\dagger}\,B^{(q-1)}_{\bs{k}}
+
\sum^{q-1}_{s=1}
\omega^{-1/3+s}\,
(A^{(s)}_{\bs{k}})^{\dagger}\,B^{(s-1)}_{\bs{k}}
\Big]
+ \textrm{H.c.}\\
&\,= t_0
(1 + \omega^{-1/6}\,
e^{-ik_1}+\omega^{-1/3}\,
e^{-i k_2})
A^{\dagger}_{\bs{k}}\,B_{\bs{k}}  + \textrm{H.c.}
\end{split}
\end{equation}

The AA-hopping reads
\begin{equation}
\begin{split}
H_{AA}
&\,=
t_1 e^{i\theta}
\sum_{\bs{k}}
\Big[
\sum^{q-2}_{s=0}
\omega^{s+1/2} A^{s+1 \dagger}_{\bs{k}} A^{s}_{\bs{k}}
+
\omega^{(q-1)+1/2} e^{-i k_2}
A^{0 \dagger}_{\bs{k}} A^{q-1}_{\bs{k}}
\Big]
\\
&\,+
t_1 e^{i\theta}
\sum_{\bs{k}}\sum^{q-1}_{s=0}
\omega^{s} e^{i k_1}
A^{s\dagger}_{\bs{k}} A^{s}_{\bs{k}}
\\
&\,+
t_1 e^{i\theta}
\sum_{\bs{k}} 
\Big[
\omega^{1} e^{-i k_1 + i k_2} 
A^{q-1 \dagger}_{\bs{k}} A^{0}_{\bs{k}} 
+ 
\sum^{q-1}_{s=1}\omega^{1-2s} e^{-i k_1} A^{s-1 \dagger}_{\bs{k}} A^{s}_{\bs{k}} 
\Big]
\\
&\,+ \textrm{H.c.}\\
&\,
=
t_1 e^{i\theta} (\omega^{1/2} e^{-i k_2} +e^{i k_1}+\omega
e^{-i k_1+i k_2})
A^{ \dagger}_{\bs{k}} A_{\bs{k}} + \textrm{H.c.}
\end{split}    
\end{equation}

The BB-hopping reads
\begin{equation}
\begin{split}
H_{BB}
&\,=
t_1 e^{i\theta}
\sum_{\bs{k}}
\Big[
\sum^{q-2}_{s=0} \omega^{5/3 + 2s} e^{i k_1} B^{s+1\dagger}_{\bs{k}} B^{s}_{\bs{k}}
+
\omega^{5/3+2(q-1)} e^{i k_1-i k_2} B^{0\dagger}_{\bs{k}} B^{q-1}_{\bs{k}}
\Big]
\\
&\,+
t_1 e^{i\theta}
\sum_{\bs{k}}
\Big[
\sum^{q-1}_{s=1} \omega^{1/6 - s} B^{s-1\dagger}_{\bs{k}} B^{s}_{\bs{k}}
+
\omega^{1/6} e^{i k_2} B^{q-1\dagger}_{\bs{k}} B^{0}_{\bs{k}}
\Big]
\\
&\,+
t_1 e^{i\theta}
\sum_{\bs{k}}
\sum^{q-1}_{s=0} \omega^{-1/3 - s} e^{-i k_1}
B^{s\dagger}_{\bs{k}} B^{s}_{\bs{k}}
\\
&\, + \textrm{H.c.}\\
&\,=
t_1 e^{i\theta}(
\omega^{5/3} e^{i k_1-i k_2}
+
\omega^{1/6} e^{i k_2}
+
\omega^{-1/3} e^{-i k_1}
)
B_{\bs{k}}^{\dagger} B_{\bs{k}}
+ \textrm{H.c.}
\end{split}
\end{equation}
Collecting previous terms leads to the 2x2 Hamiltonian is
\begin{equation}\label{2x2H}
    H[k_1,k_2,\theta,t_0,t_1] = \begin{pmatrix}
        H_{AA}& H_0\\
        H_0^*& H_{BB}
    \end{pmatrix}
    =
    h_{0}(\bs{k})\sigma_0 + \sum^{3}_{i=1}h_{i}(\bs{k})\sigma_i
    \,,
\end{equation}
with 
\begin{equation}
    \begin{split}
        &h_{0}(\bs{k}) = 2t_1\cos\theta\left(\cos k_1-\cos k_2+\cos(k_1-k_2)\right)\\
        &h_{1}(\bs{k}) = t_0(1-\cos k_1 +\cos k_2)\\
        &h_{2}(\bs{k})=t_0(-\sin k_1 +\sin k_2)\\
        &h_{3}(\bs{k}) = 2t_1\sin\theta\left(-\sin k_1-\sin k_2+\sin(k_1-k_2)\right)
    \end{split}
    \,.
\end{equation}
The energy of the lower band is directly obtained
\begin{equation}\label{dispersion}
    \varepsilon_{-}(\bs{k})
    =\frac{H_{AA}+H_{BB}}{2}-\frac{1}{2}\sqrt{(H_{AA}-H_{BB})^2+4H_0H_0^*}
    =
    h_{0}(\bs{k}) - |\bs{h}(\bs{k})|
    \,.
\end{equation}

At $k_1=-p\cdot\frac{\pi}{3},k_2=-p\cdot\frac{2\pi}{3}$, $E_{\text{low}}$ reaches the minimum, i.e.
\begin{equation}
\begin{split}
    &\partial_{k_1} \varepsilon_{-}(\bs{k}) = \partial_{k_2} \varepsilon_{-}(\bs{k}) = 0\\
    &\partial^2_{k_1} \varepsilon_{-}(\bs{k}) = \partial^2_{k_2} \varepsilon_{-}(\bs{k}) = \frac{2}{3}|t_0|+4t_1 \cos(\theta)\\
    &\partial_{k_1}\partial_{k_2} \varepsilon_{-}(\bs{k}) = -\frac{1}{3}|t_0|-2t_1 \cos(\theta)
\end{split}
\end{equation}

Expanding $\varepsilon_{-}(\bs{k})$ around the minimum,
\begin{equation}
    \varepsilon_{-}(\bs{k}) = \varepsilon^{*}_{-} +\frac{1}{2}\left(\frac{2}{3}|t_0|+4t_1 \cos(\theta)\right)\begin{pmatrix}
        k_1&k_2
    \end{pmatrix}
    \begin{pmatrix}
        1 & -1/2\\-1/2&1
    \end{pmatrix}
    \begin{pmatrix}
        k_1\\k_2
    \end{pmatrix}
\end{equation}
Considering
$\mathbf{k} = \frac{k_1}{2\pi} \vec{g}_1+\frac{k_2}{2\pi} \vec{g}_2$, with 
$\vec{g}_1 = \frac{4\pi}{3}(1/2,-\sqrt{3}/2)$ and $\vec{g}_2 = \frac{4\pi}{3}(1/2,\sqrt{3}/2)$, gives

\begin{equation}
\begin{pmatrix}
        k_x\\k_y
    \end{pmatrix} = \frac{1}{3}
    \begin{pmatrix}
        1&1\\
        -\sqrt{3}&\sqrt{3}
    \end{pmatrix}
    \begin{pmatrix}
        k_1\\k_2
    \end{pmatrix}
    \leftrightarrow
    \begin{pmatrix}
        k_1\\k_2
    \end{pmatrix} = 
    \begin{pmatrix}
        3/2&-\sqrt{3}/2\\
        3/2&\sqrt{3}/2
    \end{pmatrix}
    \begin{pmatrix}
        k_x\\k_y
    \end{pmatrix}
\end{equation}
resulting in the rotationally invariant parabolic expansion
\begin{equation}
    \begin{split}
        \varepsilon_{-}(\bs{k}) &= \varepsilon^{*}_{-} +\frac{1}{2}\left(\frac{2}{3}|t_0|+4t_1 \cos(\theta)\right)\begin{pmatrix}
        k_x&k_y
    \end{pmatrix}
    \begin{pmatrix}
        9/4 & 0\\0&9/4
    \end{pmatrix}
    \begin{pmatrix}
        k_x\\k_y
    \end{pmatrix}
    \end{split} 
\end{equation}
from which 
\begin{equation}
\varepsilon_{\text{eff}} = \frac{\hbar^2}{2m^*}
    k^2 = \left(\frac{3}{4}|t_0|+\frac{9}{2}t_1\cos(\theta)\right)k^2    
\end{equation}
yields the effective mass in Eq.\,\eqref{eq: effective mass}.

Following the same approach of Eq.\,\eqref{eq: Delta LL}, we can approximate the slopes of the energy bands in the Landau-fan  
asymptotically close to the half-filling($\nu_{\,\text{h}}=1/2$) according to
\begin{equation}\label{linreg slope}
    E_n=E^* + (n+1/2)\hbar\omega_c = E^* + (n+1/2)\frac{4}{3\sqrt{3}}\Phi_0\frac{\hbar e}{m^*}|\nu_{\,\text{h}}-1/2| \Rightarrow \mathrm{d} E_n/\mathrm{d} \delta \nu_{\,\text{h}} = \pm (n+1/2)\frac{4}{3\sqrt{3}}\Phi_0\frac{\hbar e}{m^*}
\end{equation}

\begin{figure}[!htb]
    \centering
    \includegraphics[width=0.6\linewidth]{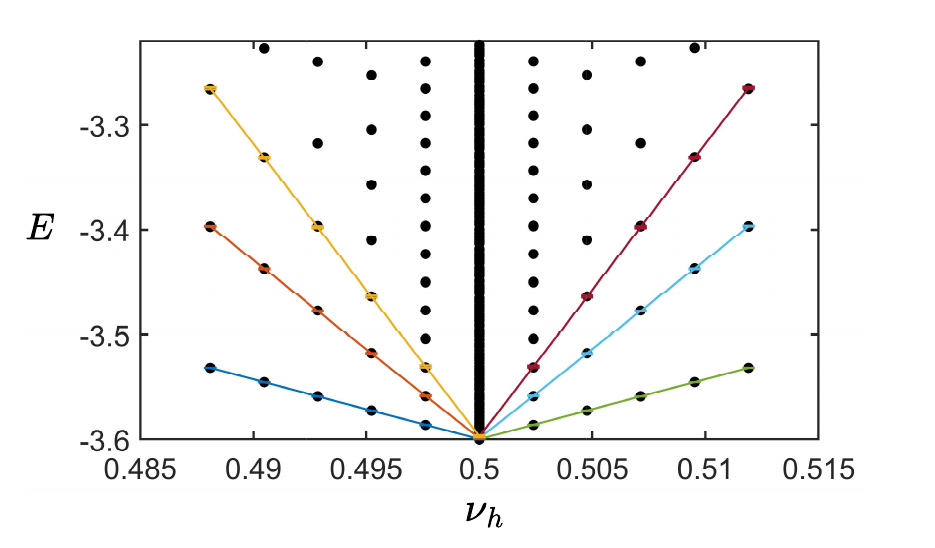}
    \caption{Linear regression of the lowest three bands in the Landau-fan at $\theta=-2\pi/3,t_0=-1,t_1=-0.2$. Solid dots are the Hofstadter spectrum, solid lines with error bars show the linear regression.}
    \label{fig: S3}
\end{figure}

To compare the slopes in Eq.\,\eqref{linreg slope} with the Hofstadter spectrum, we perform linear regression on the lowest three bands in the Hofstadter spectrum close to the half-filling at $\theta=-2\pi/3,t_0=-1,t_1=-0.2$ as shown in Fig.\,\ref{fig: S3}.

The fitting result compared with Eq.\,\eqref{linreg slope} is summarized in Table.\,\ref{tab: linear reg}, their agreement supports the Landau-fan picture accounted by the effective mass at half-filling.

\begin{table}[!htb]
    \centering
    \begin{tabular}{ |p{3cm}||p{3cm}|p{3cm}|p{3cm}|  }
 \hline
 \multicolumn{4}{|c|}{Landau-fan Slope} \\
 \hline
 \multirow{2}{3cm}{Landau level index}& \multicolumn{2}{|c|}{Linear regression} & Eq.\,\eqref{linreg slope}\\
 \cline{2-4}    & $\nu_{\,\text{h}}<1/2$ & $\nu_{\,\text{h}}>1/2$&   -\\
 \hline
 $\Delta E_0 = (0+1/2)\hbar\omega_c$&   $-5.7128\pm0.0107$  & $5.7171\pm0.0099$ &$\pm5.8042$\\
 $\Delta E_1 = (1+1/2)\hbar\omega_c$ &$-16.9621\pm0.0523$ & $16.9793\pm0.0488$&  $\pm17.4125$\\
 $\Delta E_2 = (2+1/2)\hbar\omega_c$    &$-27.8583\pm0.1342$ & $27.9012\pm0.1256$&  $\pm29.0208$\\
 \hline
\end{tabular}
    \caption{The Landau-fan slopes from linear regression and predicted by Eq.\,\eqref{linreg slope} at $\theta=-2\pi/3,t_0=-1,t_1=-0.2$.}
    \label{tab: linear reg}
\end{table}

\subsection{QPT correlated with HOVHS for $\nu_{\,\text{h}}=2/5,3/7, 6/13$}
\begin{figure}[!htb]
    \centering
    \includegraphics[width=1\linewidth]{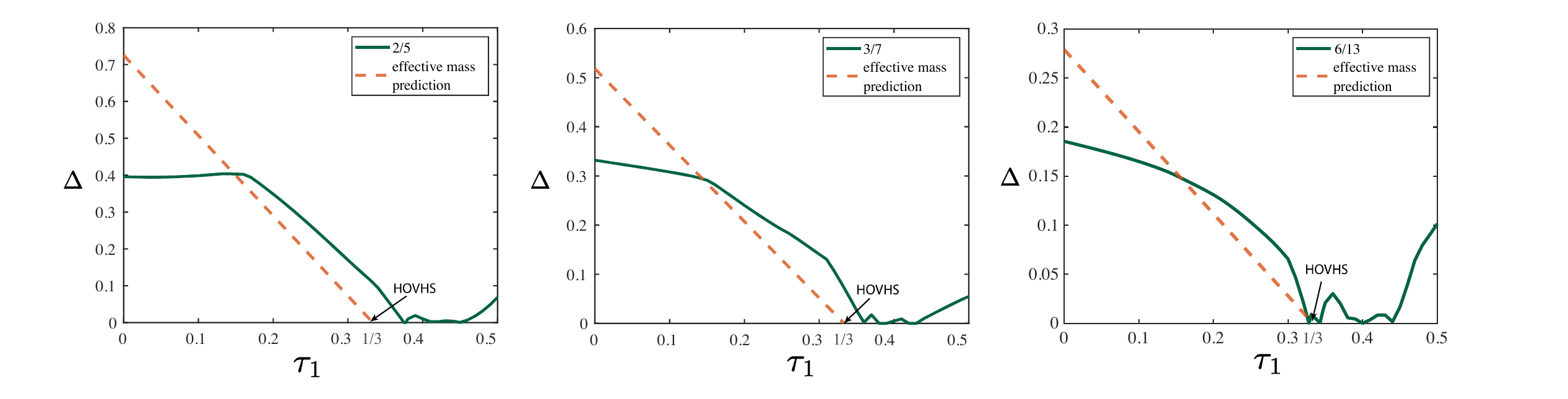}
    \caption{$\sigma=-1,\theta=-2\pi/3, t_0=-1$. The composite fermion gap versus $\tau_1$. left: $\nu_{\,\text{h}}=2/5$, middle: $\nu_{\,\text{h}}=3/7$, right: $\nu_{\,\text{h}}=6/13$. The green curve shows the gap, the orange line shows the gap predicted for each hole filling by effective mass at $\nu_{\,\text{h}}=1/2$ which vanishes at $\tau_1=1/3$ as the HOVHS emerges.}
    \label{fig: S2}
\end{figure}

Here we present three more examples of Fig.\,\ref{Fig3}(c) at $\nu_{\,\text{h}}=2/5,3/7,6/13$ (in Fig.\,\ref{fig: S2}) to show that the correlation between QPTs and HOVHS is extensive in the Jain-FCI states.

\subsection{Generalization of HOVHS to $\tau_2\neq0$}

In this section, we present generalized results for the the HOVHS for $\tau_2\neq0$.
At half-filling, $p=3,q=1$, then Eq.\,(\ref{H2 eq}) reads
\begin{equation}
    H_2 = t_2 \left(\omega^{-2/3}e^{ik_1-ik_2}+\omega^{2/3}e^{-ik_1+ik_2}+e^{-ik_1-ik_2}\right)A_{\boldsymbol{k}}^{\dagger} B_{\boldsymbol{k}}+\text { H.c. }
\end{equation}
adding the third-neighbor hopping to Eq.\,(\ref{2x2H}), we can get
\begin{equation}
    H\left[k_1, k_2, \theta, t_0, t_1, t_2\right]=\left(\begin{array}{cc}
H_{A A} & H_0+H_2 \\
H_0^*+H_2^* & H_{B B}
\end{array}\right)=h_0(\boldsymbol{k}) \sigma_0+\sum_{i=1}^3 h_i(\boldsymbol{k}) \sigma_i,
\end{equation}
where
\begin{equation}
    \begin{aligned}
& h_0(\boldsymbol{k})=2 t_1 \cos \theta\left(\cos k_1-\cos k_2+\cos \left(k_1-k_2\right)\right) \\
& h_3(\boldsymbol{k})=2 t_1 \sin \theta\left(-\sin k_1-\sin k_2+\sin \left(k_1-k_2\right)\right)\\
& h_1(\boldsymbol{k})=t_0\left(1-\cos k_1+\cos k_2\right) + t_2\left(2\cos(k_1-k_2)+\cos(k_1+k_2)\right)\\
& h_2(\boldsymbol{k})=t_0\left(-\sin k_1+\sin k_2\right)+t_2\sin(k_1+k_2)
\end{aligned}
\end{equation}

\begin{figure}[!htb]
    \centering
    \includegraphics[width=0.9\linewidth]{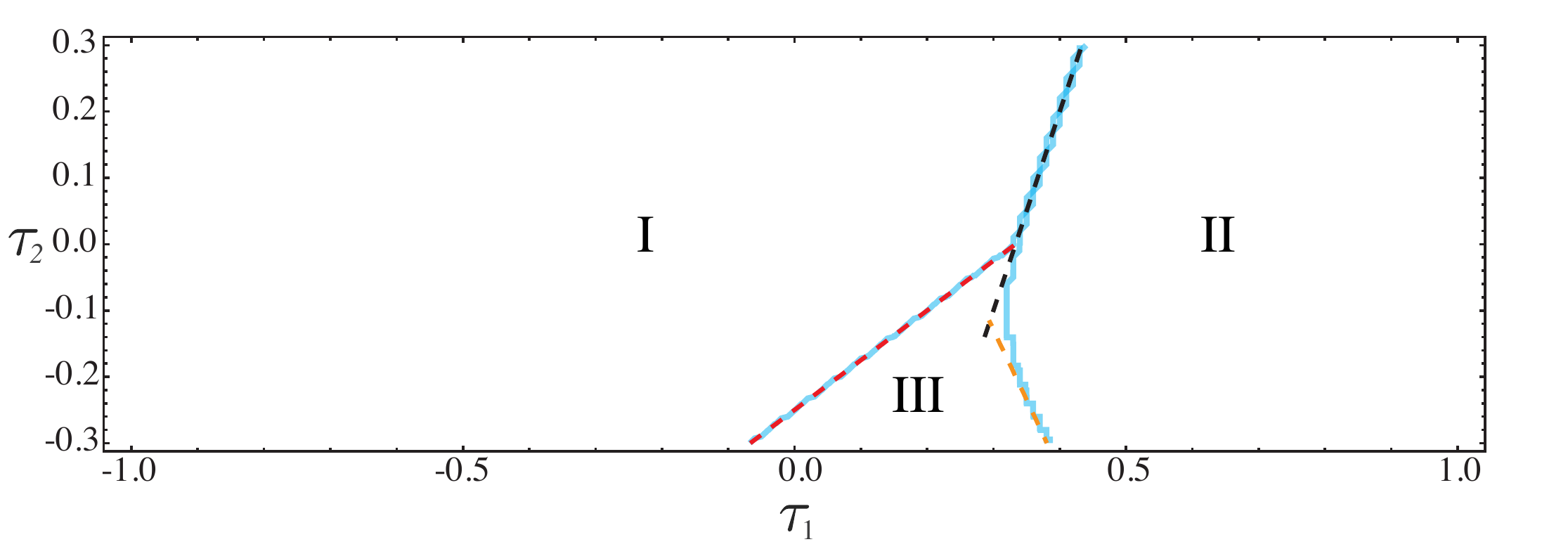}
    \caption{Phase diagram of global minimum in Eq.\,(\ref{dispersion}). In phase I, the global minimum is fixed at
    $k_1=\pi,k_2=2\pi$. In phase II, the global minimum is fixed at
    $k_1=5/3\pi,k_2=4/3\pi$. In phase III, the minimum forms a Mexican-hat dispersion. Dashed lines mark the theoretic boundaries between each phase, the blue curves show the boundaries by numerical calculations.
    }
    \label{FigS4}
\end{figure}

And the dispersion follows Eq.\,(\ref{dispersion}), for simplicity, we fix $\theta=-2\pi/3$ and $t_0=-1$.

There are three phases as shown in Fig.\,\ref{FigS4}. In phase I, the global minimum of the lower band is fixed at $k_1=\pi,k_2=2\pi$, and the parabolic expansion is
\begin{equation}
\varepsilon_{\text{eff}} = \frac{\hbar^2}{2m^*}
    k^2 = \left(\frac{3}{4}-\frac{9}{4}\tau_1+3\tau_2\right)k^2    
\end{equation}
and in phase II, the global minimum of the lower band is fixed at $k_1=5/3\pi,k_2=4/3\pi$, and the parabolic expansion is
\begin{equation}
\varepsilon_{\text{eff}} = \frac{\hbar^2}{2m^*}
    k^2 = \left(
    \frac{9}{2}\tau_1-\frac{1}{\tau_1}(-\frac{1}{2}+\tau_2)^2
    \right)k^2    
\end{equation}
while in phase III, the minimum of $\varepsilon_{-}(\boldsymbol{k})$ forms a Mexican-hat dispersion.

At the interface between phase I and II (black dashed line in Fig.\,\ref{FigS4}), there is no HOVHS observed, while the transition is described by
\begin{equation}
    3\tau_1-\tau_2-1=0
\end{equation}

At the interface between phase I and III (red dashed line in Fig.\,\ref{FigS4}), the HOVHS occurs when effective mass $m^*$ diverges, which is characterized by  
\begin{equation}
    \frac{3}{4}-\frac{9}{4}\tau_1+3\tau_2=0
\end{equation}

At the interface between phase II and III (orange dashed line in Fig.\,\ref{FigS4}), the HOVHS is observed when effective mass $m^*$ diverges, which is characterized by  
\begin{equation}
    \frac{9}{2}\tau_1-\frac{1}{\tau_1}(-\frac{1}{2}+\tau_2)^2=0
\end{equation}

\end{document}